\documentclass[pre,preprint,unsortedaddress]{revtex4-1} 
\usepackage{amsmath}  % needed for \tfrac, \bmatrix, etc.
\usepackage{amsfonts} % needed for bold Greek, Fraktur, and blackboard bold
\usepackage{graphicx} % needed for figures
\usepackage{enumitem} % needed for enumeration
\usepackage{xcolor} % needed for coloured text
\usepackage[utf8]{inputenc} %needed for Erdos double accute accents
\usepackage{tikz} % for inset
\usepackage{csquotes}
\usepackage[subrefformat=parens,labelformat=parens]{subfig}

\begin{document}

% Be sure to use the \title, \author, \affiliation, and \abstract macros
% to format your title page.  Don't use lower-level macros to  manually
% adjust the fonts and centering.

\title{Superdiffusion on complex networks: the role of shortcuts and long-range interactions}
% In a long title you can use \\ to force a line break at a certain location.

\author{Alfonso Allen-Perkins}
\email{alfonso.allen@hotmail.com}
\affiliation{Complex System Group, Universidad Polit\'ecnica de Madrid, 28040-Madrid, Spain.\\
Instituto de F\'isica, Universidade Federal da Bahia, 40210-210 Salvador, Brazil.}%

\author{Alfredo Blanco Serrano}
\email{alfredoblancoserrano@gmail.com}
\affiliation{Instituto de F\'isica, Universidade Federal da Bahia, 40210-210 Salvador, Brazil.}%

\author{Thiago Albuquerque de Assis}
\email{thiagoaa@ufba.br}
\affiliation{Complex System Group, Universidad Polit\'ecnica de Madrid, 28040-Madrid, Spain.\\
Instituto de F\'isica, Universidade Federal da Bahia, 40210-210 Salvador, Brazil.}%

\author{Juan Manuel Pastor}
\email{juanmanuel.pastor@upm.es}
\affiliation{Complex System Group, Universidad Polit\'ecnica de Madrid, 28040-Madrid, Spain.\\
E.T.S.I.A.A.B, Universidad Polit\'ecnica de Madrid, Avd. Puerta de Hierro 4, 28040-Madrid, Spain.}%

%\author{Ernesto Estrada}
%\email{ernesto.estrada@strath.ac.uk}
%%\noaffiliation
%\affiliation{Department of Mathematics and Statistics, University of Strathclyde, Glasgow, G1 1XQ, U.K.}%

\author{Roberto F. S. Andrade}
\email{randrade@ufba.br}
\affiliation{Instituto de F\'isica, Universidade Federal da Bahia, 40210-210 Salvador, Brazil.}%

\date{\today}

\begin{abstract}
This work addresses the superdiffusive motion of a discrete time random walker on ordered discrete substrates and complex networks with the presence of long-range interactions (LRIs). In ordered regular lattices, where LRIs have a clear geometrical meaning, their presence allow for hoppings between more distant sites, yet with a smaller probability. In such cases, it is found that LRIs do not affect the dependency of the mean square displacement (MSD) traveled by the walker: exact analytical results for the the cycle graph within the Markov chain framework shows that MSD follows the same linearly increasing behavior with time when LRIs are absent, independently of the strength of LRI. This contrasts with the superdiffusive scenario in complex networks. When they have very short diameter ($\sim \log N$), the analysis of the time dependency of MSD becomes quite difficult, as it saturates very quickly even when LRIs are absent. The presence of a faster than linearly increasing growth phase can be noticed, but it can hardly be measured with precision. This effect is sidestepped on small-world Newman-Watts (NW) networks, where the network diameter can be controlled by the number of new links (shortcuts) that are added to the cycle graph. The time duration $t_f$ of the superdiffusive regime and the power law exponent can be adequately evaluated by numerical methods. They depend on the number of nodes and shortcuts, as well as the strength of LRIs. Although the later causes a strong reduction in $t_f$ when shortcuts are present, their presence by itself is not sufficient to trigger a superdiffusive behavior.
\end{abstract}

\maketitle 

\section{Introduction}

In real-world complex systems many dynamical processes occur in discrete spaces. They include diffusion, synchronization of agents, epidemic spreading, and many other processes taking place in ecological, social, economic as well as infrastructural and technological systems \cite{estrada2012structure}. These systems can be represented by networks, in which the nodes represent the constituent entities and the edges represent the interactions between such entities. Dynamical processes on these networks usually adopt a nearest-neighbor strategy of transferring ``information'' from one node to another. %Then, one of the main modeling scenarios used for the study of dynamical processes on networks is
Here, random walk models have widespread use in the analysis of diffusion of information and navigability on these networks, as well as in the exploration of their structures to detect their fine-grained organization \cite{Masuda17,noh2004random,klafter2011first}.

Currently, it is well-documented that there are dynamical processes both in continuum and discrete spaces which do not follow this ``nearest-neighbor paradigm'' being under the influence of non-local, long-range distance interactions. For instance, self-diffusive processes of atoms and molecules adsorbed on metals display significant contributions due to jumps spanning more distant atoms %in the metallic surface. %For instance, when studying the diffusion of two large organic molecules on the Cu(110) surface it was observed that long jumps play a dominating role in the diffusionof the two organic molecules, with root-mean-square jump length as large as $3.9$ and $6.8$ lattice spacing} \cite{schunack2002long}. Since then
%The role of long-jumps in ad-atoms and molecules diffusing
on metallic surfaces, a phenomenon that has been both theoretically and experimentally confirmed in many different systems \cite{schunack2002long,yu2013single,ala2002collective,guantes01}.
In the continuum space, the use of L\'evy flights is very frequent to model a large variety of processes in which long-range jumps occurs together with short-range ones, % The most typical examples are those related to the diffusive 
as in the motion of species in a given environment \cite{lomholt2008levy,humphries2010environmental,song2010modelling,rhee2011levy}.%. Another useful approach to study such processes is to consider
The effect of long range interactions can also be conveniently described by fractionary differential equations \cite{bakunin11} or non-linear diffusion equations \cite{tsalis09}. More recently, another approach has emerged as an alternative to study dynamical processes on discrete spaces, which combine nearest- and non-nearest-neighbors hops in an elegant mathematical way. This approach is based on the so-called $d$-path Laplacian operators, which represent a natural generalization of the Laplacian operator on graphs \cite{estrada12,estrada17b}. It has been employed to analyze consensus spreading or synchronization, among other phenomena \cite{estrada13,estrada17,estrada17c}.

In the current work, we investigate the behavior of the mean square displacement (MSD) of a discrete time random walker as a function of time on finite networks, looking for the possible existence of super-diffusive dynamics. As it was proven in \cite{estrada17c} that the generalized diffusion equation using the Mellin-transformed $d-$path Laplacians can produce superdiffusive processes on infinite path graphs (linear chains), it is natural to inquire whether a similar scenario holds also for finite complex networks. In particular, we use the model with long-range jumps described in \cite{estrada17b}, to better understand the effects of the network topology and long distance jumps. Thus, this work advances under the previous results on hitting and commute times presented in \cite{estrada17b}.

Most used network types, like those generated within the Erd\H{o}s-R\'enyi \cite{Erdos1959} and Barab\' asi-Albert frameworks \cite{Barabasi1999}, are characterized by short diameter, which favors diffusion saturation in a very short time scales. Because of this, and also by the possibility of developing an exact analytical treatment of the diffusion properties, we first focus our attention to the ordered, large diameter cycle graphs. Our analytical developments provide advances over the results described in \cite{montroll65} for the one-dimensional lattice with periodic boundary conditions. Then, we compare these results with those obtained for the small-world Newman-Watts (NW) model \cite{Newman99}, where randomness arises by adding some new connections (shortcuts) in a controlled way. As we will show, the interplay between less probable long distance jumps and extra shortcuts leads to a rich diffusive pattern.

The paper is organized as follows. In Sec. \ref{RW-time}, we present the formalism used to describe discrete time random walks on complex networks with long-range interactions (LRIs), as well as the methodology employed to estimate the MSD on these systems. Numerical results supporting the analytical ones for the time evolution of the MSD on cycle graphs with or without LRIs are shown in Secs. \ref{Results-cons-LRIS}, whereas the findings for NW networks with or without LRIs are discussed in \ref{Results-SW-cons-LRIS}. Our conclusions are summarized in Sec. \ref{Con}. Finally, a detailed derivation of the mathematical expressions used can be consulted in the appendices.

%%%%%%%%%%%%%%%%%%%%%%%%%%%%%%%%%%%%%%%%%%%%%%%%%%%%%
%%%%%%%%%%%%%%%%%%%%%%%%%%%%%%%%%%%%%%%%%%%%%%%%%%%%%
%%%%%%%%%%%%%%%%%%%%%%%%%%%%%%%%%%%%%%%%%%%%%%%%%%%%%

\section{Discrete time random walks on networks with long-range interactions and the estimation of MSD}
\label{RW-time}

Let $G=(V,E)$ be a simple, undirected graph or network without self-loops. The usual \textit{discrete time random walk} on $G$ is a random sequence of vertices generated as follows: given a starting vertex $i$ we select a neighbor $j$ randomly, and move to this neighbor. Then we select a neighbor $k$ of $j$ randomly, and move to it, and so on \cite{Masuda17,aldous02,lovasz93}. %In this section we introduce the definitions and methodology 
The characterization of this system amounts to calculate the probability of finding a random walker at a given node at time $t$, when the random walker is initially located at node $i$. The model we investigate here includes the probability of long distance jumps, which can be thought as resulting from weaker LRIs in the network. Thus, to include this effect in the evaluation of  the desired probability distribution, we follow the methodology presented in \cite{estrada17b}.

Let $d_{ij}$ be the shortest path distance, that is, the number of edges in the shortest path connecting the nodes $i$ and $j$, and let $d_\mathrm{max}$ be the graph diameter, that is, the maximum shortest path distance in the graph. Let us now define the $d-$path adjacency matrix \cite{estrada17b} (or neighborhood adjacency matrix of order $d$ \cite{andrade06}), denoted by $\mathbf{A}_{d}$, of a connected graph of $N$ nodes as the square, symmetric, $N \times N$ matrix whose entries are:

\begin{equation}
\mathbf{A}_{d}\left(i,j\right)=\left\{ \begin{array}{r}
1\\
0
\end{array}\right.\begin{array}{l}
\textnormal{if \ensuremath{d_{ij}=d} }\\
\textnormal{otherwise}
\end{array},
\end{equation}

\noindent where $d \leq d_\mathrm{max}$.

Let us now consider the transformed $d-$path adjacency matrices of $G$ given by:

\begin{equation}
\hat{\mathbf{A}} ^\tau =\left\{ \begin{array}{r}
\sum _{d =1}^{d_\mathrm{max}}  d^{-s} \mathbf{A}_d\\
\mathbf{A}_1+\sum_{d =2}^{d_\mathrm{max}} e^{-\lambda d}\mathbf{A}_d
\end{array}\right.\begin{array}{l}
\textnormal{if \ensuremath{\tau=\mathrm{Mellin}}}\\
\textnormal{if \ensuremath{\tau=\mathrm{Laplace}}}
\end{array},
\label{adj_transf_1}
\end{equation}

\noindent where $\tau$ indicates the type of transformation, $s \geq 0$ and $\lambda \geq 0$ are constant values. In the case of $s=-1$ and $\tau=\mathrm{Mellin}$, $\hat{\mathbf{A}} ^\tau$ is equal to the neighborhood matrix as defined in \cite{andrade06}.

Following \cite{Masuda17,estrada17b,zhang13}, let us define the strength of a given node $i$ of a transformed $d-$path graph as:

\begin{equation}
\hat{s}^\tau \left ( i \right )=\left ( \hat{\mathbf{A}}^{\tau} \vec{1} \right )_i
\end{equation}

\noindent where $\vec{x}$ is an all-x vector. Consequently, the probability that a particle staying at node $i$ hops to the node $j$ is given by:

\begin{equation}
P^\tau \left ( i,j \right )= \frac{\hat{\mathbf{A}}^{\tau} \left (i,j \right )}{\hat{s}^\tau \left ( i \right )}.
\end{equation}

Let us denote by $\hat{ \mathbf{\mathcal{S}}}^\tau$ the diagonal matrix with elements $\hat{ \mathbf{\mathcal{S}}}^\tau \left ( i,i \right )=\hat{s}^\tau \left ( i \right )$ and let us define the \textit{transition matrix} for the random walk as ${ \mathbf{\mathcal{P}}}= \left ( \hat{ \mathbf{\mathcal{S}}}^\tau \right ) ^{-1}\hat{\mathbf{A}}^\tau$. According to this definition, it is possible to see that ${ \mathbf{\mathcal{P}}}$ is a stochastic matrix, and that the vector containing the probability of finding a random walker at a given node of the graph at time $t+1$ is given by

\begin{equation}
\vec{p}_{t+1}= { \mathbf{\mathcal{P}}} ^T \vec{p}_{t},
\end{equation}

\noindent where $\mathbf{X}^T$ stands for the transpose of matrix $\mathbf{X}$.

The vector $\vec{p}_{t}$ depends on the initial position of the random walker. We denote by $\vec{p}_{t,i}$ the vector containing the probability of finding a random walker at a given node of the graph at time $t$, when the random walker is initially located at node $i$. Therefore, it is possible to write the following expression:

\begin{equation}
\vec{p}_{t,i}=\mathbf{\mathcal{P}}_{t-1}^T \cdots \mathbf{\mathcal{P}}_{0}^T \vec{p}_{0,i}= \left (\mathbf{\mathcal{P}}^T \right )^t \vec{p}_{0,i},
\label{time_evolution_prob}
\end{equation}

\noindent where $\left ( \vec{p}_{0,i} \right )_j=1$ if $i=j$, and $0$ otherwise.

%%%%%%%%%%%%%%%%%%%%%%%%%%%%%%%%%%%%%%%%%%%%%%%%%%%%%%%%%
%%%%%%%%%%%%%%%%%%%%%%%%%%%%%%%%%%%%%%%%%%%%%%%%%%%%%%%%%
%%%%%%%%%%%%%%%%%%%%%%%%%%%%%%%%%%%%%%%%%%%%%%%%%%%%%%%%%

%\section{MSD on networks with time-dependent long-range interactions}
%\label{MSD}

The MSD is a measure of the distance between the position of a walker at a time $t$, $x(t)$, and a reference position, $x_0$. In most cases, this quantity is described by an expression of the form:

\begin{equation}
\left \langle \left ( x(t)-x_0 \right )^2 \right \rangle = \left \langle r^2(t) \right \rangle \sim t^\gamma,
\end{equation}

\noindent where the value of the parameter $\gamma$ classifies the type of diffusion
into normal diffusion ($\gamma=1$), subdiffusion  ($\gamma<1$), or superdiffusion ($\gamma>1$). MSD \cite{almaas03,gallos04} is one of the most common way to analyze stochastic data. However, in order to characterize diffusion, additional complementary measures are usually required, e.g., first passage observables. In fact, random-walk with LRIs in \cite{estrada17b} were characterized by hitting and commute times, instead of MSD. Other aspects the problem, including usual  network diffusion and network synchronization \cite{estrada13,estrada17} are often discussed based on the eigenvalues of the Laplacian matrices. For the type of results we discuss here, MSD is essential to provide a clear cut way to characterize the time dependence.

The MSD has been studied on simple networks (i.e., unweighted graphs without self-loops) by means of random walkers \cite{almaas03,gallos04}. In this case, this quantity is a measure of the distance $r$ covered by a typical walker after performing $t$ steps.

Given $G=(V,E)$ and an initial condition $\vec{p}_{0,i}$, we calculate the MSD of the random walker to the origin (i.e., the node $i$), at each time step, $r^2(t,i)$, as follows:

\begin{equation}
r^2(t,i)=\sum_{j=1}^{N} \left ( d_{i,j} \right )^2 \left ( \vec{p}_{t,i} \right )_j.
\label{MSD_una_condicion_new}
\end{equation}

To obtain numerical estimates for the MSD, we average over all the different initial positions of the walker:

\begin{equation}
\mathrm{MSD}\equiv \left \langle r^2(t) \right \rangle = \frac{1}{N}\sum_{i=1}^N r^2(t,i)= \frac{1}{N}\sum_{i=1}^N \sum_{j=1}^{N}\left ( d_{i,j} \right )^2 \left ( \vec{p}_{t,i} \right )_j.
\label{eq_MSD_multpl_new}
\end{equation}

As can be observed, %the assesment of MSD on 
the MSD definition, first introduced to describe single step, nearest neighbor walks in ordered structures and adopted in complex networks in the sequence, can be easily extended to networks with long-range interactions. According to Eqs. \ref{time_evolution_prob} and \ref{eq_MSD_multpl_new}, $\left \langle r^2(t) \right \rangle$ depends on various factors, namely: the discrete time-step $t$, the d-path transformation used $\tau$ (Mellin or Laplace), the weight of the d-path transformation ($s$ or $\lambda$), and the topology of $G$. For the sake of simplicity, in this work we only consider the case of $\tau=\mathrm{Mellin}$.

%%%%%%%%%%%%%%%%%%%%%%%%%%%%%%%%%%%%%%%%%%%%%%%%%%%%%%%%%%%
%%%%%%%%%%%%%%%%%%%%%%%%%%%%%%%%%%%%%%%%%%%%%%%%%%%%%%%%%%%

\section{Time evolution of the MSD on cycle graphs with LRIs}
\label{Results-cons-LRIS}

As advanced in the Introduction, in order to sidestep the short diffusive regime frequently appearing in complex networks with small diameter, we focus our attention to the special class of NW networks, which can be obtained within the NW model by starting from an ordered cycle graph and randomly adding a small amount of new extra links. This way, we can follow how the exact analytical results for the ordered structure can be compared to the results obtained from numerical simulations for the cycle graph itself, and for the derived NW complex networks. Although the Watts-Strogatz (WS) model \cite{watts1998collective}, based on rewiring a fraction of the original short range connections, could also be used for this purpose, our choice is mainly motivated to avoid the emergence of isolated clusters, which is not forbidden within the WS algorithm.

So let us initially consider the case of a cycle graph with $N$ nodes. When $s=\infty$ (i.e., the cycle graph has no LRIs) and $N$ is an odd number, the following exact expression for the MSD can be derived,

\begin{eqnarray}
\left \langle r^2(t) \right \rangle= \frac{N^2-1}{12} +\sum_{k=2}^{(N+1)/2}  \cos^t\left ( \theta_k \right )\frac{(-1)^{k+1}}{\sin \left ( \frac{\theta _k}{2} \right )} \cot\left ( \frac{\theta _k}{2} \right ),
\label{eq_MSD_s_inf1}
\end{eqnarray}

\noindent where $\theta_k \equiv \frac{2\pi}{N}\left ( k-1 \right )$ (see Sec. A of Appendix C for the derivation of Eq.~\ref{eq_MSD_s_inf1}).

When the Mellin transformation is used and $s \leq \infty$, the above expression can be generalized for any value value of $s$ as 

\begin{eqnarray}
\left \langle r^2(t) \right \rangle = \frac{N^2-1}{12} +\sum_{k=2}^{(N+1)/2}  
\frac{(-1)^{k+1}}{\sin \left ( \frac{\theta _k}{2} \right )} \cot\left ( \frac{\theta _k}{2} \right )
\left ( \frac{1}{H_{0,\frac{N-1}{2}} ^s} \sum_{d=1}^{(N-1)/2}\frac{\cos\left ( \theta_k d\right )}{d^s} \right )^t.
\label{eq_MSD_s_cons}
\end{eqnarray}

\noindent Here $H_{c,n}^{(m)}$ is the generalized harmonic number for nonnegative $n$, complex order $m$ and complex offset $c$ \cite{kronenburg12}, defined as

\begin{eqnarray}
H_{c,n}^{(m)}=\sum_{k=1}^{n}\frac{1}{(c+k)^m}
\end{eqnarray}

\noindent (see Sec. B of Appendix C for the derivation of Eq.~\ref{eq_MSD_s_cons}). %As can be seen in Fig \ref{ejemplo_ajuste_0}, there is a complete agreement between the results of Eqs. \ref{eq_MSD_multpl_new} and \ref{eq_MSD_s_cons}.

%%%%%%%%%%%%%%%%%%%%%%%%%%%%%%%%%%%%%%%%%%%%%%%
%\begin{figure}[h!]
%\centering
%\includegraphics[width=0.5\textwidth]{fig0.eps}
%\caption{Time evolution of MSD for a cycle graph with LRIs, when $N=101$ nodes and $s=3$. The red circles indicate the numerical results obtained according to Eq.~\ref{eq_MSD_multpl_new}, whereas the curve shows those of  Eq.~\ref{eq_MSD_s_cons}.}
%\label{ejemplo_ajuste_0}
%\end{figure}
%%%%%%%%%%%%%%%%%%%%%%%%%%%%%%%

From Eq.~\ref{eq_MSD_s_cons}, it is possible to derive initial and asymptotic time regimes for $\left \langle r^2(t) \right \rangle$ on cycle graphs with LRIs, which can be written as

%The numeric estimation of Eq.~\ref{eq_MSD_multpl_new} on cycle graphs with LRIs shows that it is possible to write the following approximation for Eq.~\ref{eq_MSD_s_cons}:

\begin{eqnarray}
\left \langle r^2(t) \right \rangle = \left\{ \begin{array}{r}
\left \langle r^2(1) \right \rangle t\\
\left \langle r^2 \right \rangle_{\mathrm{sat}}
\end{array}\right.\begin{array}{l}
\textnormal{if \ensuremath{1\leq t\ll t_{\mathrm{x}}}}\\
\textnormal{if \ensuremath{t\gg  t_{\mathrm{x}}}}
\end{array},
\label{approx_s_cons}
\end{eqnarray}

\noindent where

\begin{eqnarray}
\left \langle r^2 \right \rangle_{\mathrm{sat}} =  \frac{N^2-1}{12},
\label{MSD_sat_def}
\end{eqnarray}

\begin{eqnarray}
 \left \langle r^2(1) \right \rangle \equiv \left \langle r^2(t=1) \right \rangle  =  \frac{\sum _{k=1}^{(N-1)/2} \left ( \frac{N+1}{2} -k\right )^{2-s}}{\sum _{k=1}^{(N-1)/2} k^{-s}}=\left ( -1 \right )^{s-2}\frac{H_{-\frac{N+1}{2}, \frac{N-1}{2} }^{(s-2)}}{H_{0, \frac{N-1}{2} }^{(s)}},
\label{epsilon_def}
\end{eqnarray}

\noindent %\textcolor{blue}{is the value of MSD at $t=1$ (derived from Eq.~\ref{eq_MSD_multpl_new} for cycle graphs),}
and

\begin{eqnarray}
t_{\mathrm{x}}=\frac{\left \langle r^2 \right \rangle_{\mathrm{sat}}}{\left \langle r^2(1) \right \rangle},
\label{t_sat_def}
\end{eqnarray}

\noindent represents the crossover time between the growth and saturation regimes (see Sec. F of Appendix C for the derivation of Eq.~\ref{MSD_sat_def}).

%Note that in case of $s=\infty$, $\left \langle r^2(1) \right \rangle=1$ (see Fig.~\subref*{MSD_ini_cycle}). As can be observed in Fig.~\ref{ejemplo_ajuste}, Eq.~\ref{approx_s_cons} is a fairly good approximation for the time evolution of MSD when LRIs are considered. According to Eq.~\ref{approx_s_cons}, the MSD on finite cycle graphs with LRIs exhibits a normal diffusion ($\gamma=1$) before saturation takes place (i.e., for $t\ll t_x$), as illustrated in Fig.~\subref*{various_MSD_cons}, for several values of $s$ and $N$. 

\begin{figure}[h!]
\centering
\subfloat[]{\label{fig1a}\includegraphics[width=0.5\textwidth]{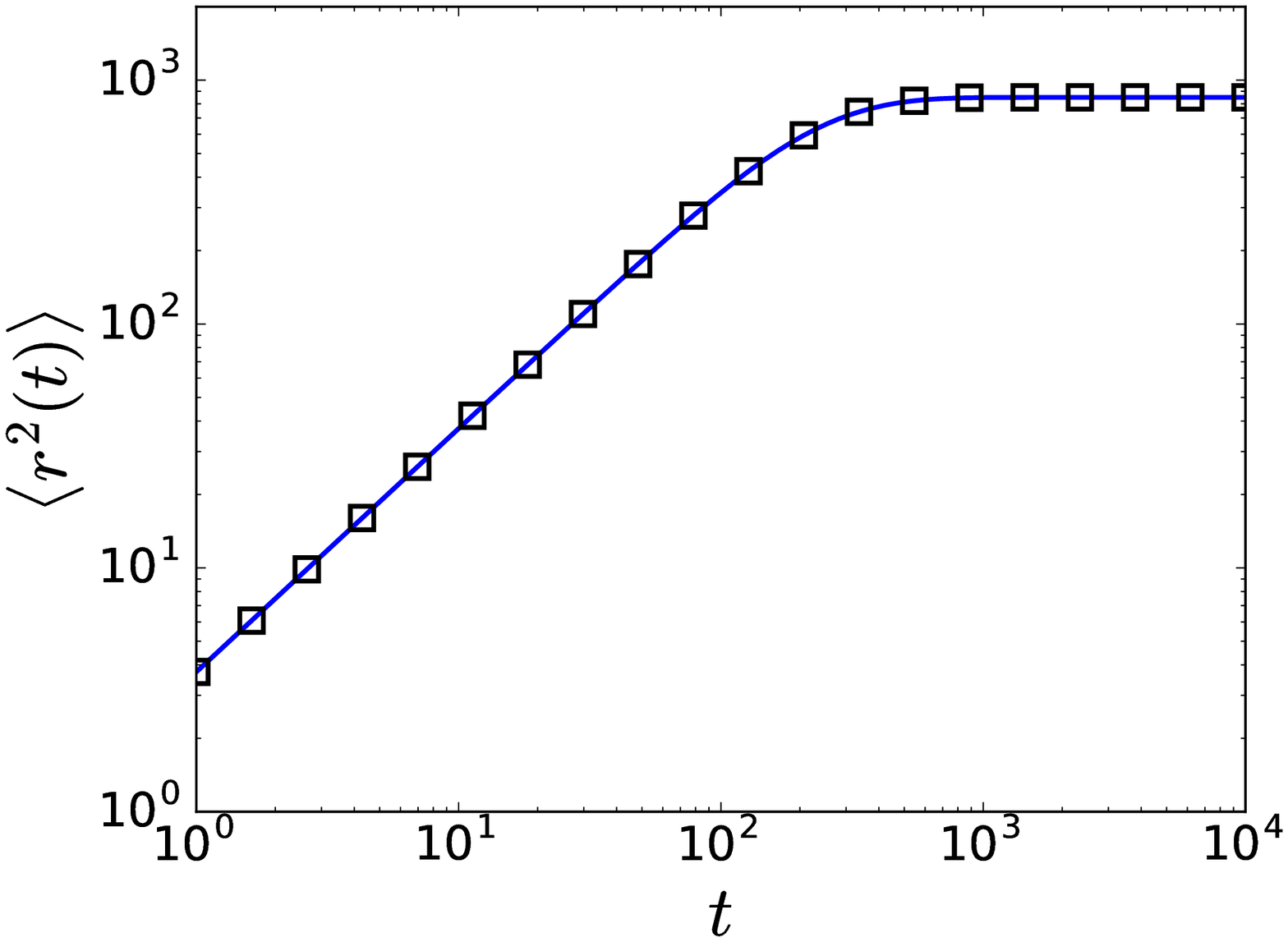}}
\subfloat[]{\label{fig1b}\includegraphics[width=0.5\textwidth]{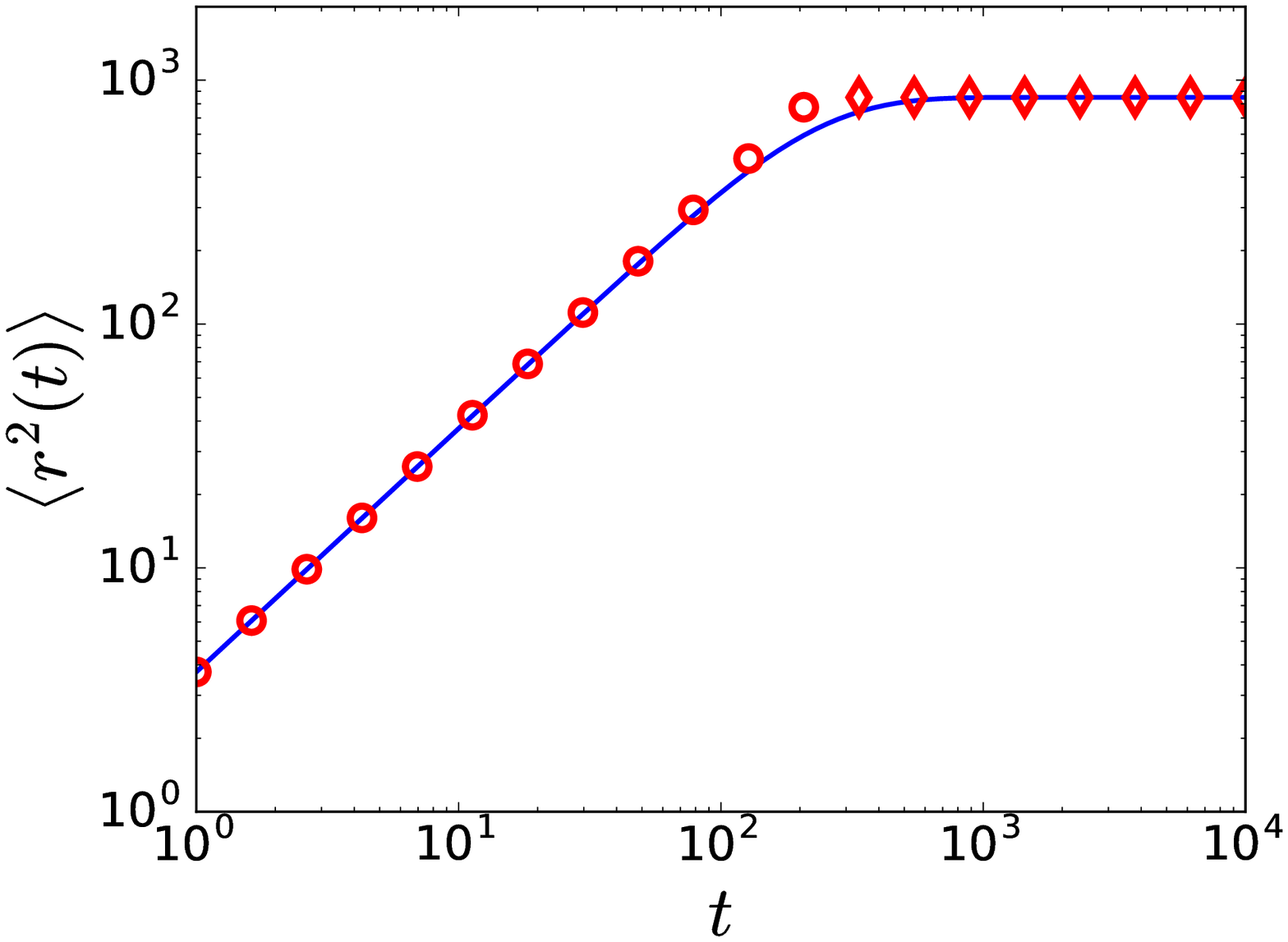}}
\caption{Time evolution of MSD for a cycle graph with LRIs, when $N=101$ nodes and $s=3$. The blue curve indicate the numerical results obtained according to Eq.~\ref{eq_MSD_s_cons}. (a) Time evolution of MSD according to Eq.~\ref{eq_MSD_multpl_new} (black squares). (b) Approximate time evolution of MSD according to Eq.\ref{approx_s_cons} (red circles and red diamonds for, respectively, the initial and asymptotic regimes) .}
\label{ejemplo_ajuste}
\end{figure}

\begin{figure}[h!]
\centering
\subfloat[]{\label{various_MSD_cons}\includegraphics[width=0.5\textwidth]{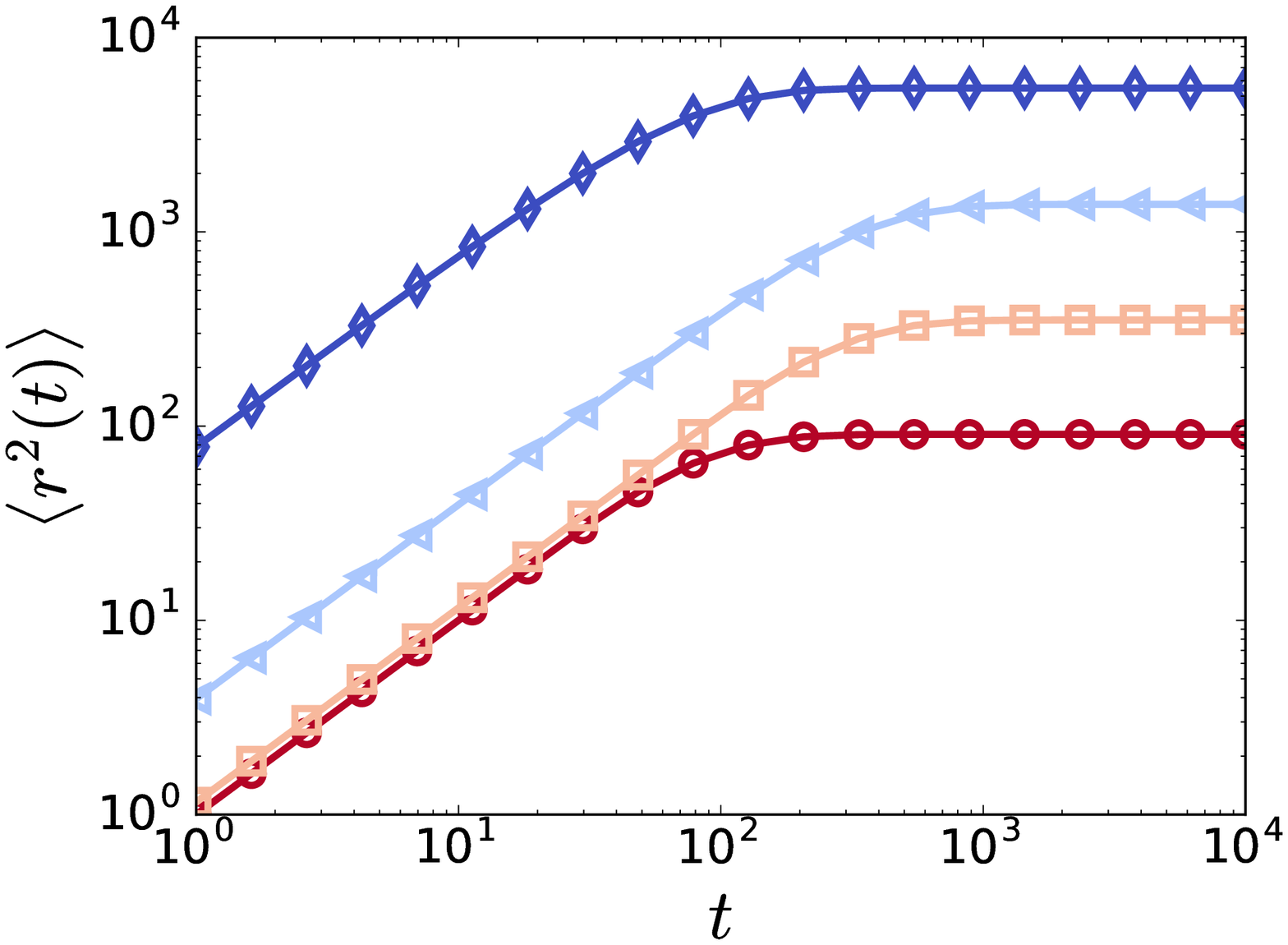}}
\subfloat[]{\label{various_MSD_collapse_cons}\includegraphics[width=0.5\textwidth]{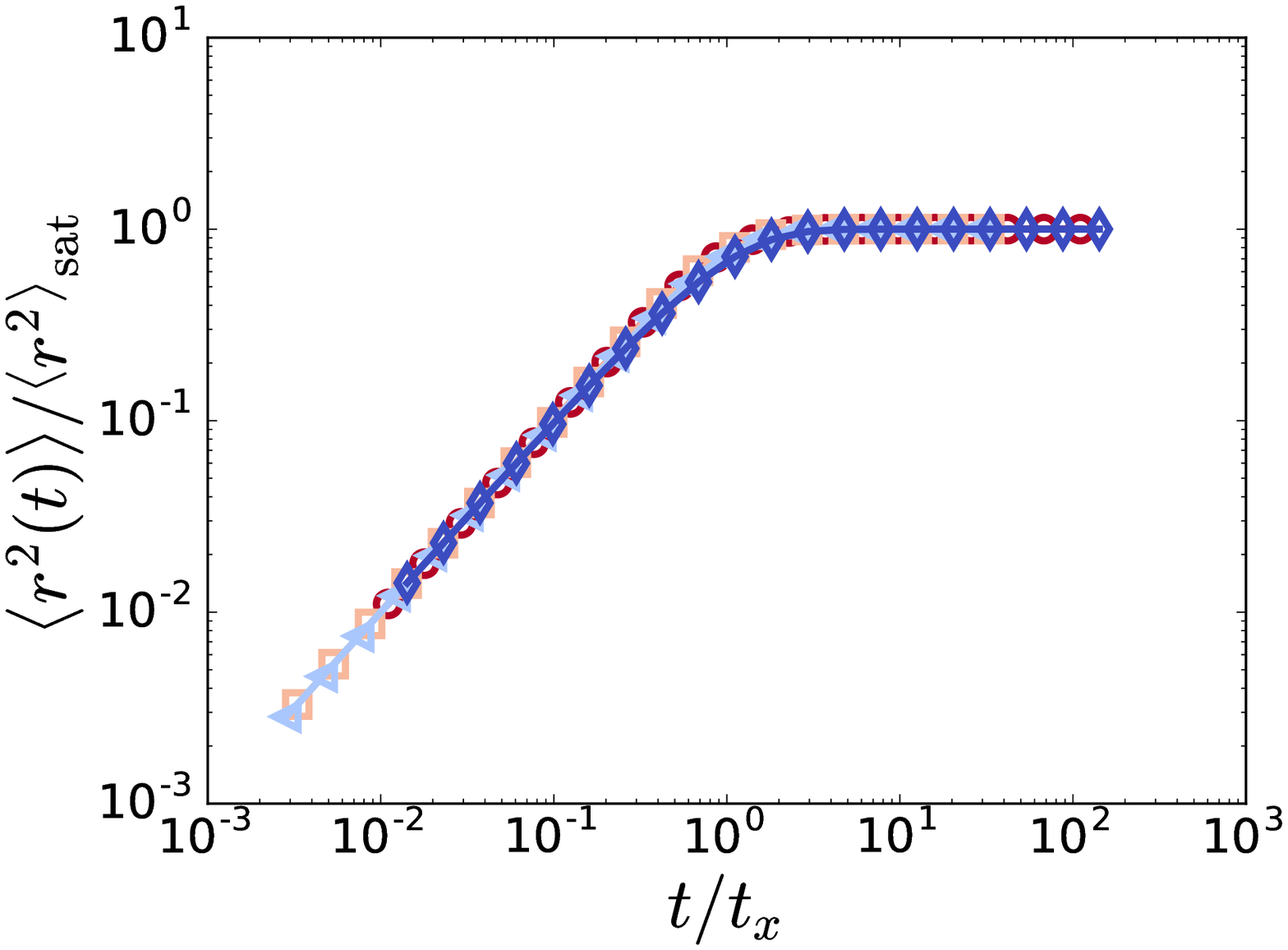}}
\caption{(a) Time evolution of the MSD for a cycle graph with LRIs, when $N=33$ and $s=\infty$ (red circles), $N=65$ and $s=5$ (orange squares), $N=129$ and $s=3$ (cyan triangles), and $N=257$ and $s=2$ (blue diamonds). The symbols indicate the result of Eq.~\ref{eq_MSD_multpl_new}, whereas the curves show those of Eq.~\ref{eq_MSD_s_cons}. (b) Collapse of the prior curves after rescaling with Eqs. \ref{MSD_sat_def}-\ref{t_sat_def}.}
\label{collapse_cons}
\end{figure}

Fig.~\ref{ejemplo_ajuste} illustrates the time evolution of $\left \langle r^2(t) \right \rangle$ for cycle graphs with the presence of LRIs. Panel (a) shows the complete agreement between the results of Eqs.~\ref{eq_MSD_multpl_new} and \ref{eq_MSD_s_cons}, while in (b) we show that Eq.~\ref{approx_s_cons} is a fairly good approximation for the limiting regimes in the previous expressions. The essential features are reproduced also in Fig.~\ref{collapse_cons}, for five different values of $N$ and $s$. The exact and approximate expressions indicate that the MSD on finite cycle graphs with LRIs exhibits normal diffusion ($\gamma=1$) before saturation takes place (i.e., for $t \ll t_{\mathrm{x}}$). Moreover, using the analytical expressions for $\left \langle r^2 \right \rangle_{\mathrm{sat}}$, $\left \langle r^2(1) \right \rangle$ and $t_{\mathrm{x}}$ as a function of $s$ and $N$ in Eqs. \ref{MSD_sat_def}, \ref{epsilon_def} and \ref{t_sat_def}, it is possible to rescale all curves and obtain a single universal function describing MSD with LRIs on cycle graphs, as shown in Fig.~\subref*{various_MSD_collapse_cons}.

%%%%%%%%%%%%%%%%%%%%%%%%%%%%%%%%%%%%%%%%%%%%%%%
%%%%%%%%%%%%%%%%%%%%%%%%%%%%%%%

%%%%%%%%%%%%%%%%%%%%%%%%%%%%%%%%%%%%%%%%%%%%%%%%%%
%%%%%%%%%%%%%%%%%%%%%%%%%%%%%%%%%%%%%%%%%%%%%%%%%%%

%The normal diffusion of the discrete-time random walkers found in these finite cycle graphs constitutes a novel result. 

\begin{figure}[h!]
\centering
\subfloat[]{\label{cyclos_s_2_varios}\includegraphics[width=0.5\textwidth]{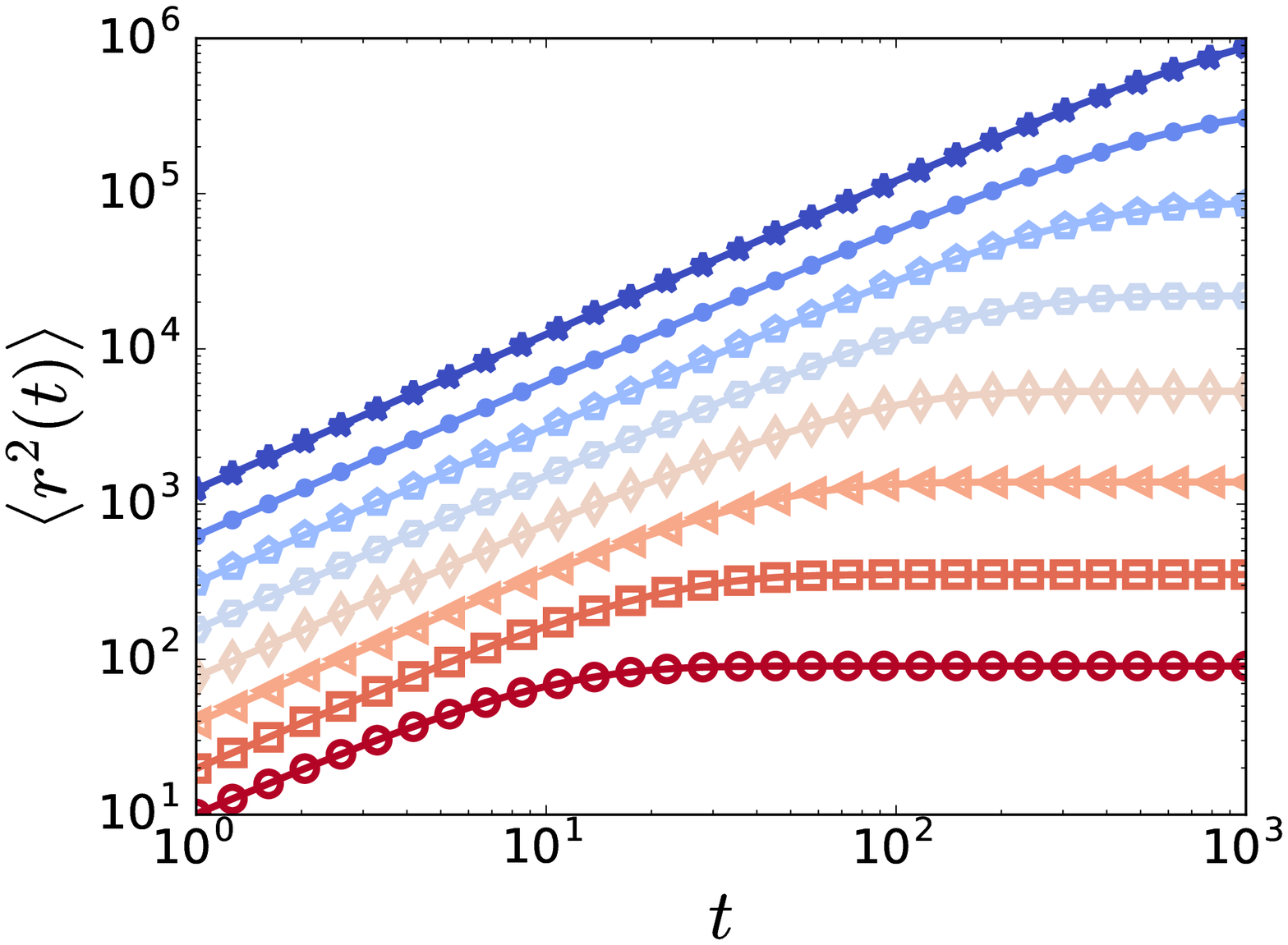}}
\subfloat[]{\label{coll_cyclos_s_2_varios}\includegraphics[width=0.5\textwidth]{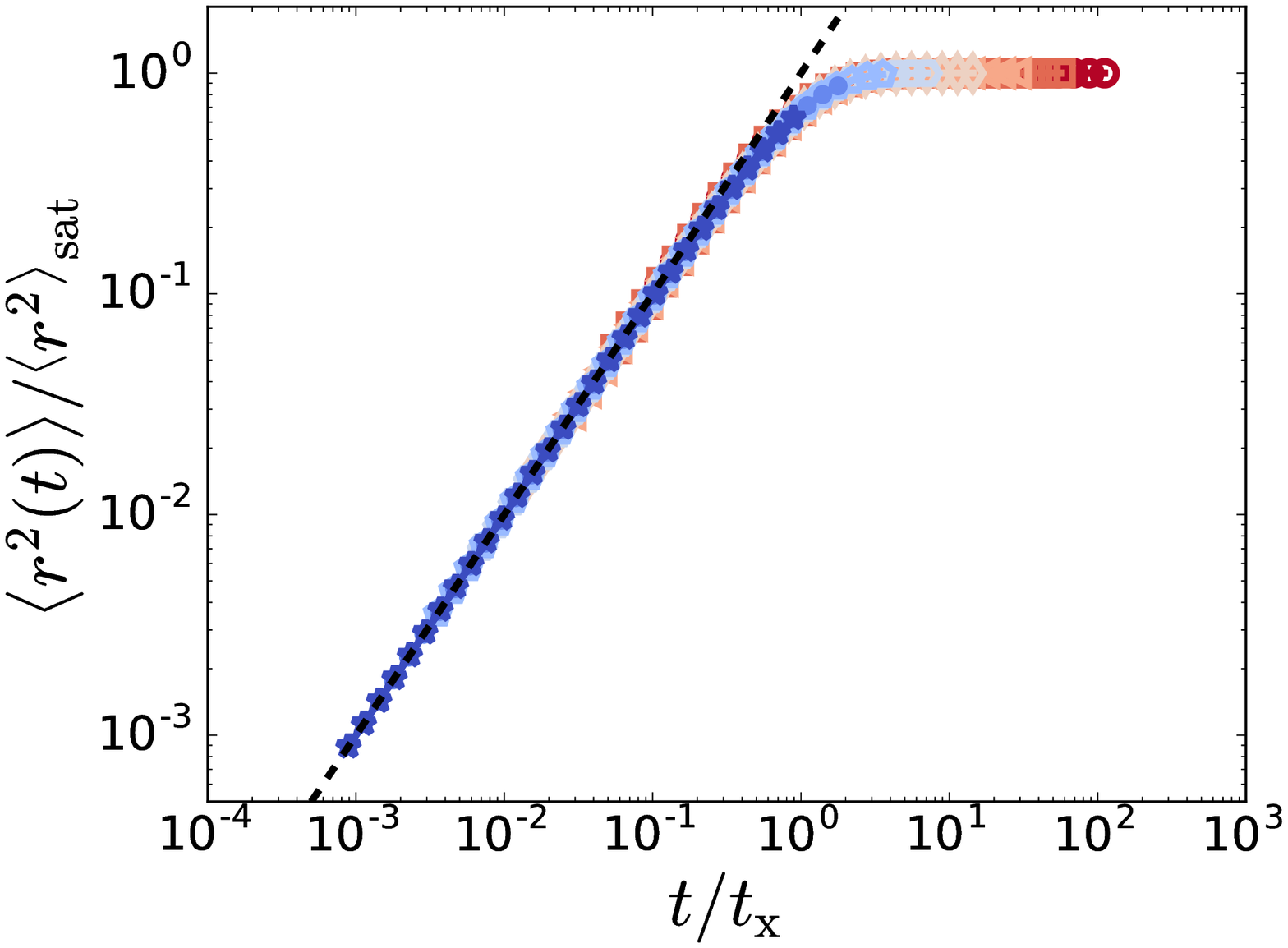}}
\caption{(a) Time evolution of the MSD for a cycle graph with LRIs, when $s=2$ (Eq.~\ref{eq_MSD_s_cons}): $N=33$ (circles), $N=65$ (squares), $N=129$ (triangles), $N=257$ (diamonds), $N=513$ (hexagons), $N=1025$ (pentagons), $N=2049$ (dots), and $N=4097$ (asteriks). (b) Collapse of the prior curves after rescaling. The black dashed line is a guide for the eye to locate a normal diffusion.}
\label{collapse_s_2}
\end{figure}

Recently, it has been reported that the generalized diffusion equation using the Mellin-transformed $d-$path Laplacians produces superdiffusive processes on infinite paths (and continuous time), as $t\rightarrow \infty$, when $1 <s <3$ \cite{estrada17c}. A detailed numerical investigation based on the Eq.~\ref{eq_MSD_s_cons} for $s=2$ and quite large values of $N$ is shown in Fig.~\ref{collapse_s_2}. The curves are in complete agreement with our findings in Fig.~\ref{collapse_cons} for several values of $s$. They exhibit the usual diffusion behavior, indicating that the results in \cite{estrada17c} do not apply for discrete-time random walkers. 

%%%%%%%%%%%%%%%%%%%%%%%%%%%%%%%%%%%%%%%%%%%%%%%%%%
%%%%%%%%%%%%%%%%%%%%%%%%%%%%%%%%%%%%%%%%%%%%%%%%%%%

We call the attention that the used $d-$path model can also be put in correspondence with the truncated L\'evy flights (TLF) \cite{Mantegna}. This model, which has some similarities with $d-$path models analyzed here and in \cite{estrada17c}, also considers random walks on the continuous infinite linear substrate, but with an $\alpha-$Lévy distribution of jump sizes ($0<\alpha<2$) that is truncated to avoid jumps larger than a parameter $\ell$. The probability of return time $R(t)$ decreases with time according to a power-law with exponent 0.5, for the long time regime which depends on $\ell$, and with exponent $1/\alpha$ in the $\ell$ transient time. 

\begin{figure}[h!]
\centering
\includegraphics[width=0.6\textwidth]{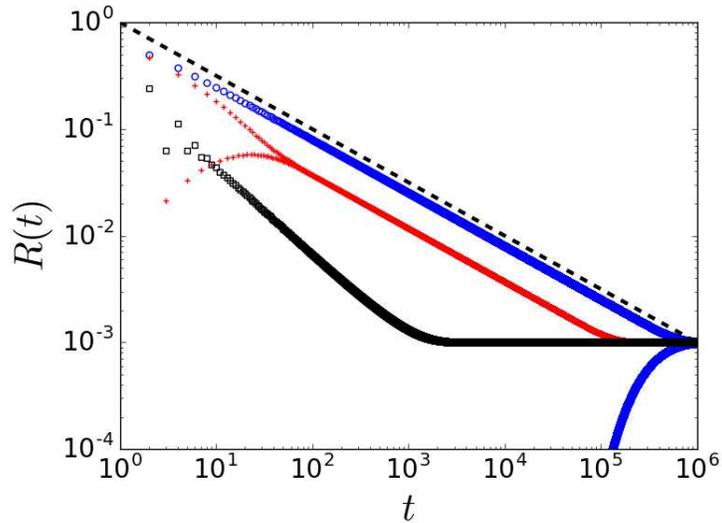}
\caption{Time evolution of probability of return time $R(t)$ for a cycle graph with LRIs, when $N=1001$ nodes and $s=\infty$ (blue circles), $s=5$ (red crosses) and $s=2.2$ (black squares). The black dashed line represents $t^{-1/2}$.}
\label{evol_R}
\end{figure}

Numerical calculations for our model indicate that $R$ reproduces the quoted TLF dependency: for $s \leftrightarrow \alpha+1>3$, $R$ decreases with a power-law exponent $\sim 0.5$, until it reaches the equiprobable value $1/N$ (see Fig.~\ref{evol_R}). On the other hand, when $s<3$, we find that the decaying exponent becomes very close to $1/(s-1)$. Because of the similar behavior, we conjecture that the TFL might also not be able to lead super-diffusion.

%%%%%%%%%%%%%%%%%%%%%%%%%%%%%%%%%%%%%%%%%%%%%%%
%%%%%%%%%%%%%%%%%%%%%%%%%%%%%%%

%. In fact, it is possible to obtain an universal curve for MSD on cycle graphs. \textcolor{blue}{In Fig.~\subref*{various_MSD_collapse_cons}, we show the collapse of the MSD curves obtained for various cycle graphs. To do so, we analytically evaluate the values of $\left \langle r^2 \right \rangle_{\mathrm{sat}}$, $\left \langle r^2(1) \right \rangle$ and $t_{\mathrm{x}}$ of each series, using Eqs. \ref{MSD_sat_def}, \ref{epsilon_def} and \ref{t_sat_def}.} As can be observed, the slope of the universal curve is equal to 1, for $t \ll t_{\mathrm{x}}$. %During the saturation regime (i.e., $t \approx t_{\mathrm{x}}$), a transient subdiffusive behavior appears.

\begin{figure}[h!]
\centering
\subfloat[]{\label{MSD_ini_cycle}\includegraphics[width=0.5\textwidth]{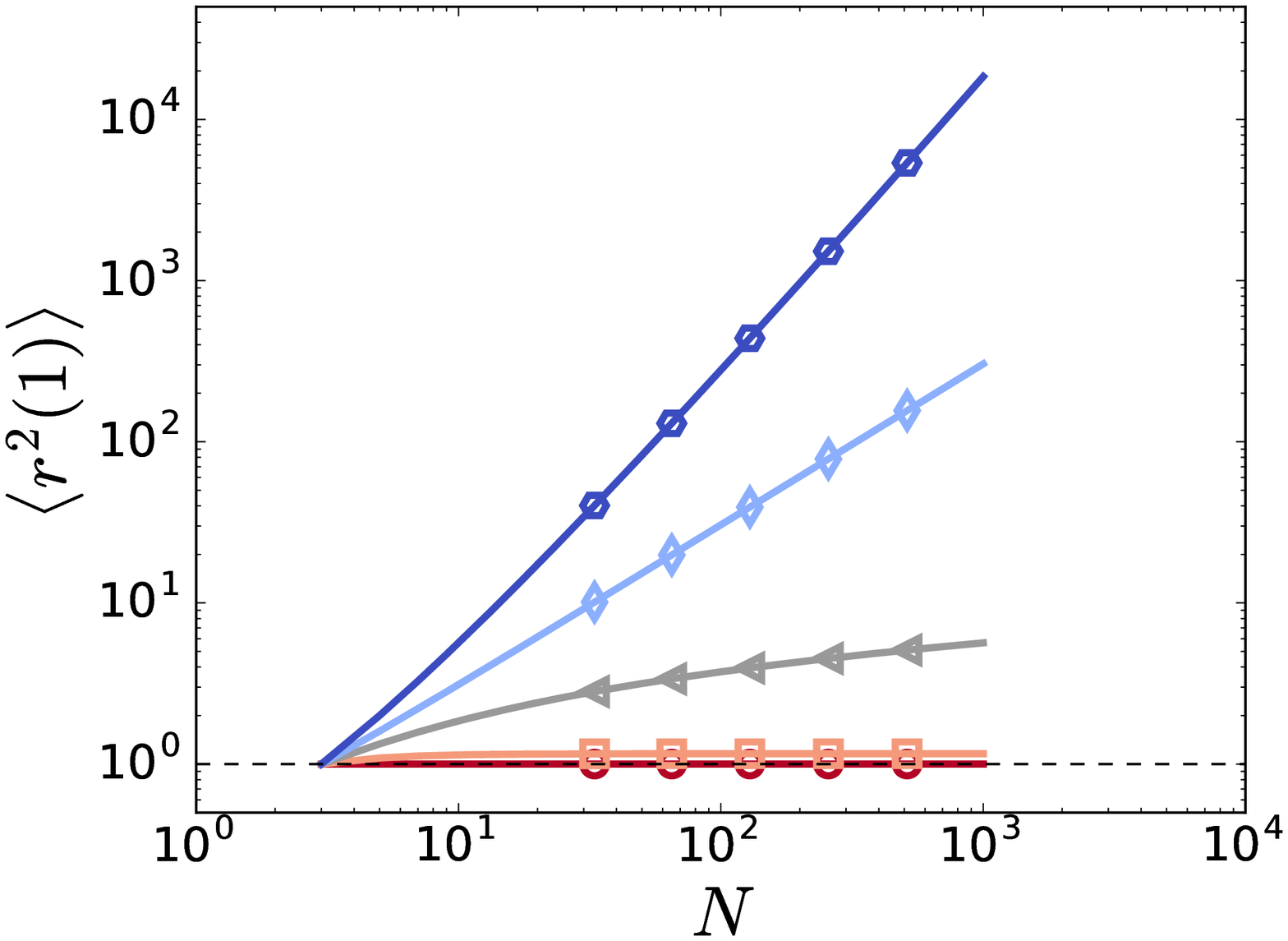}}
\subfloat[]{\label{t_sat_cycle}\includegraphics[width=0.5\textwidth]{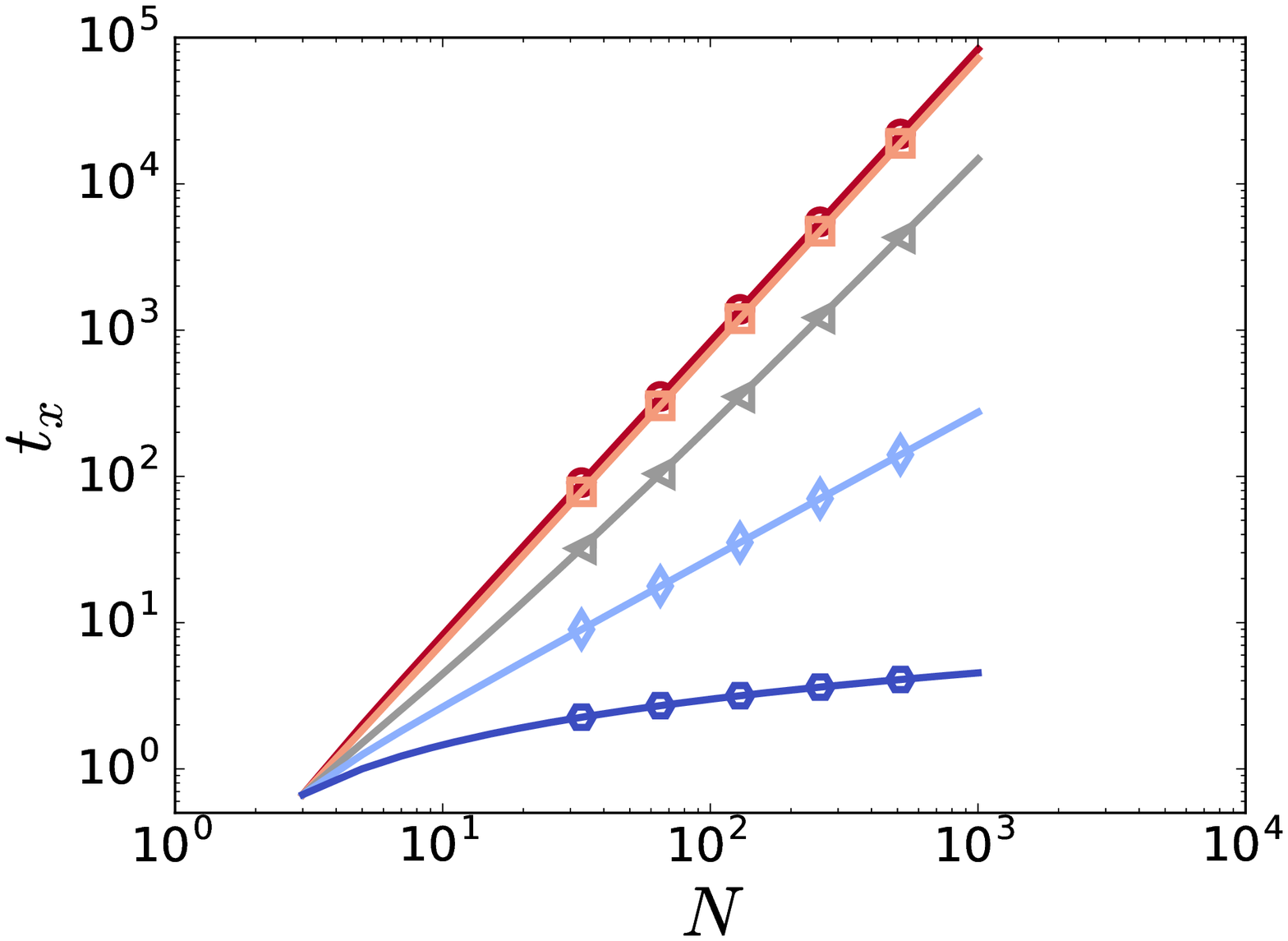}}
\caption{Dependence of $\left \langle r^2(1) \right \rangle$ (a) and $t_{\mathrm{x}}$ (b) on $s$ and $N$ for cycle graphs, when the Mellin transformation is used. The solid lines represent the values obtained analytically in Eqs. \ref{MSD_sat_def}-\ref{t_sat_def} for different values of s and all the odd values of $N \in [3,1001]$. The symbols indicate the results obtained from the simulations for $N = 33$, 65, 129, 257 and 513, and correspond to $s = \infty$ (red circles), $s = 5$ (orange squares), $s = 3$ (grey triangles), $s = 2$ (cyan diamonds), and $s = 1$ (blue hexagons).}
%\textcolor{blue}{(a) Dependence of MSD at $t=1$ on $s$ and $N$ for cycle graphs, when the Mellin transformation is used. The black dashed line represents $\left \langle r^2(1) \right \rangle = 1$. (b) Dependence of the crossover time $t_{\mathrm{x}}$ on $s$ and $N$ for cycle graphs, when the Mellin transformation is used.The markers indicate the results obtained from the simulations for $N=33$, 65, 129, 257 and 513, respectively, and the solid lines represent the values obtained analytically from Eq.~\ref{t_sat_def} for different values of $s$ and all the odd values of $N$ between 3 and 1001: $s=\infty$ (red circles), $s=5$ (orange squares), $s=3$ (grey triangles), $s=2$ (cyan diamonds) and $s=1$ (blue hexagons).}}
\label{teor_cycle}
\end{figure}

In Fig.~\subref*{MSD_ini_cycle}, we exhibit the dependence of $\left \langle r^2(1) \right \rangle$ on $s$ and $N$ (odd numbers in the interval $[3,1001]$) for a cycle graphs. As can be seen, for a given value of $N$, the larger the value of $s$, the smaller $\left \langle r^2(1) \right \rangle$. It is possible to observe that, for $N \gtrsim 10$ and $s\gtrsim 5$, $\left \langle r^2(1) \right \rangle \approx1$, and, in the case of $s = \infty$, $\left \langle r^2(1) \right \rangle = 1$, for $N\geq 3$. On the other hand, for a given value of $s < 5$, the larger the value of $N$, the larger $\left \langle r^2(1) \right \rangle$. %It is worth mentioning that, the curves of constant $s$ are concave for $s>2$, and convex for $0\leq s \lesssim 2$.

Finally, in Fig.~\subref*{t_sat_cycle}, we exhibit the dependence of the crossover time $t_{\mathrm{x}}$ on $s$ and $N$ for a cycle graphs. As can be seen, for a given value of $N$, the larger the value of $s$, the larger $t_{\mathrm{x}}$. It is possible to observe that, for a given value of $s$, the larger the system size $N$, the larger $t_{\mathrm{x}}$. Using Eqs. \ref{MSD_sat_def}-\ref{t_sat_def}, we obtain that, when $s \gtrsim 5$, the crossover time can be approximated by $t_{\mathrm{x}} \propto N^2$. However, this approximate dependence on $N$ changes when we increase the strength of LRIs (i.e when $0<s<5$). For example, in the case of $s=1$, the result is $t_{\mathrm{x}} \propto1.53\log_{10}N$.

%%%%%%%%%%%%%%%%%%%%%%%%%%%%5
%%%%%%%%%%%%%%%%%%%%%%%%%%%%%%%

%%%%%%%%%%%%%%%%%%%%%%%%%%%%%%%%%%%%%%%%%%%%%%%%%%%%%%%%%%%%%%%%%%%%%%%%%%%%%%%%%%%%%%%%%%%%%%%%%%%%%%%%%%%%%%%%%%%%%%%%%%%%%%%%%%%%%%%%%%%%%%%%%%%%%%%%%%%%%%%%%%%%%%%%%%%%%%%%%%%%%%%%%%%%%%%%%%%%%%%%%%%%%%%%%%%%%%%%%%%%%%%%%%%%%%%%%%%%%%%%%%%%%%%%%%%%%%%%%%%%%%%%%%%%%%%%%%%%%%%%%%%%%%%%%%%%%%%%%%%%%%%%%%%%%%%%%%%%%
%%%%%%%%%%%%%%%%%%%%%%%%%%%%5

\section{Time evolution of the MSD on NW networks with LRIs}
\label{Results-SW-cons-LRIS}

\begin{figure}[h!]
\centering
\subfloat[]{\label{new_WS_0p0}\includegraphics[width=0.3\textwidth]{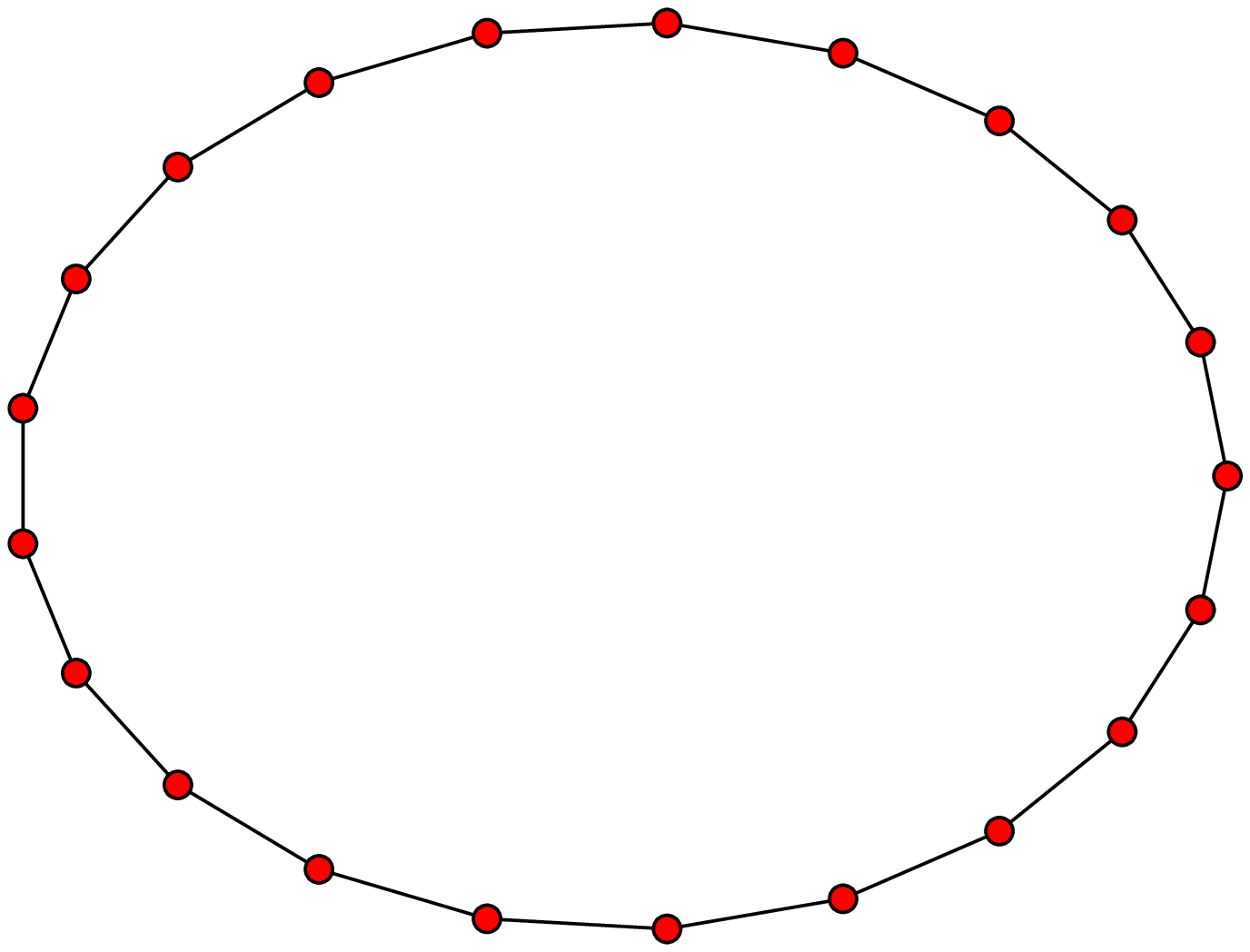}}
\subfloat[]{\label{new_WS_0p1}\includegraphics[width=0.3\textwidth]{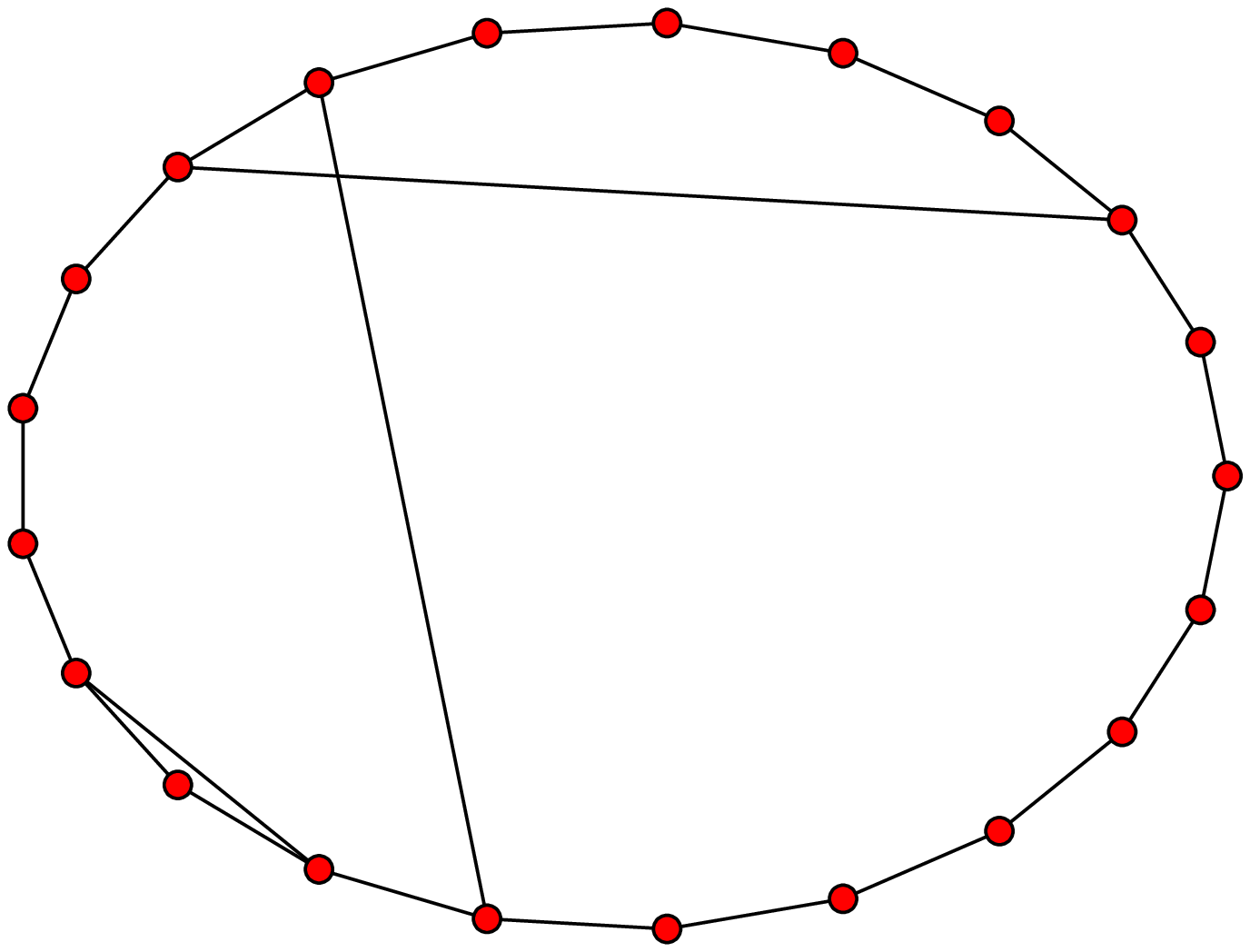}}
\subfloat[]{\label{new_WS_0p5}\includegraphics[width=0.3\textwidth]{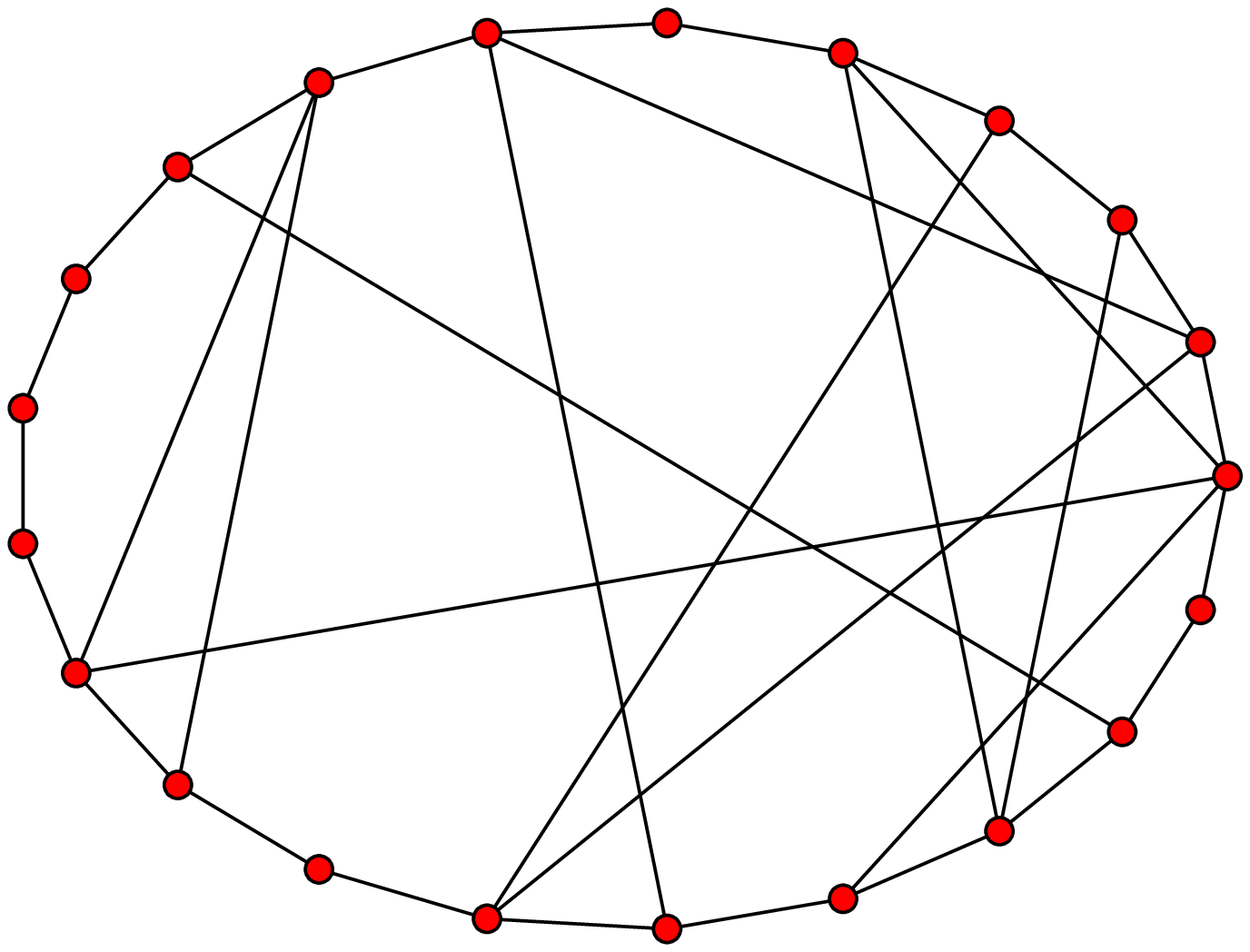}}
\caption{Schematic illustration of the NW small-world network model \cite{Newman99}, when each node is connected to 2 nearest neighbors in a ring topology, and $N=21$. (a) $p=0.0$ (cycle graph, $d_\mathrm{max}=10$). (b) $p=0.1$ ($d_\mathrm{max}=7$). (c) $p=0.5$ ($d_\mathrm{max}=4$).}
\label{ejemplos_shortcuts}
\end{figure}

In this section we numerically explore the behavior of MSD on NW networks \cite{Newman99}, which are obtained by adding some extra connections (shortcuts) to a cycle graphs in the following way:
%In this section, using Eq.~\ref{eq_MSD_multpl_new} and the NW networks \cite{Newman99}, we explore numerically the effect of adding shortcuts to cycle graphs. According to the model in \cite{Newman99}, we first create a ring over $N$ nodes. Then, the shortcuts are created by adding new edges as follows:
 for each edge $i-j$ (in the underlying \enquote{N-ring}) a new link $i-k$ (with a randomly-chosen node $k$) is added with probability $p$. In Fig.~\ref{ejemplos_shortcuts}, we show the results obtained with this model, for various values of $p$. As expected, the larger the value of $p$, the smaller the diameter of the resulting network, $d_\mathrm{max}$.

%%%%%%%%%%%%%%%%%%%%%%%%%%%%%%%%%%%%
%%%%%%%%%%%%%%%%%%%%%%%%%%%%%%%%%%%%%%%%%%%%%%%%%%%%%%

\begin{figure}[h!]
\centering
%\subfloat[]{\label{WS_MSD_N_21}\includegraphics[width=0.5\textwidth]{WS_MSD_N_21}}
\subfloat[]{
   \begin{tikzpicture}
        \node[anchor=south west,inner sep=0] (image) at (0,0) {\includegraphics[width=0.5\textwidth]{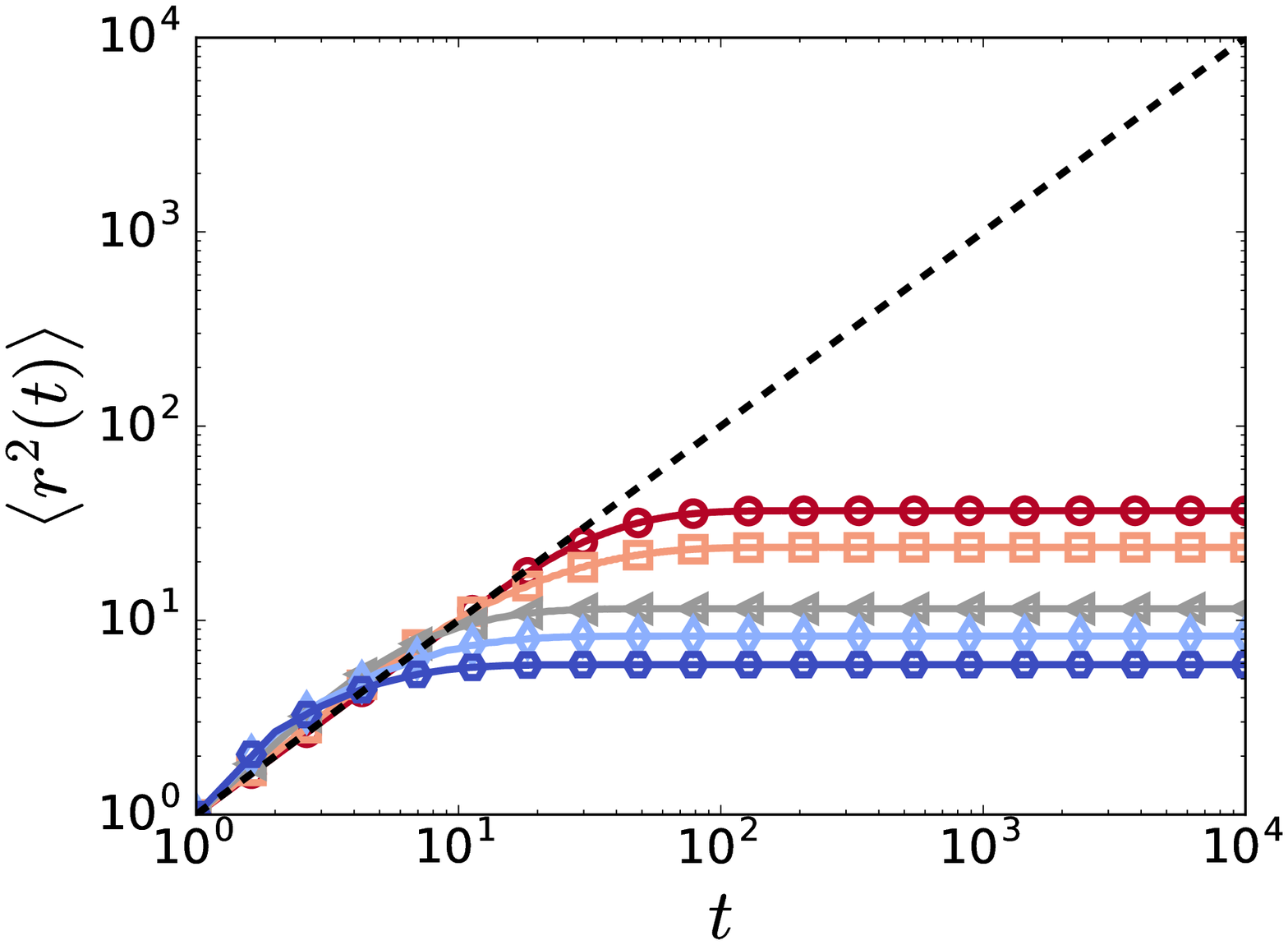}};
        \begin{scope}[x={(image.south east)},y={(image.north west)}]
            %\draw[help lines,xstep=.1,ystep=.1] (0,0) grid (1,1);
            %\foreach \x in {0,1,...,9} { \node [anchor=north] at (\x/10,0) {0.\x}; }
            %\foreach \y in {0,1,...,9} { \node [anchor=east] at (0,\y/10) {0.\y}; }
            \node[anchor=south west,inner sep=0] (image) at (0.18,0.55) {\includegraphics[width=0.18\textwidth]{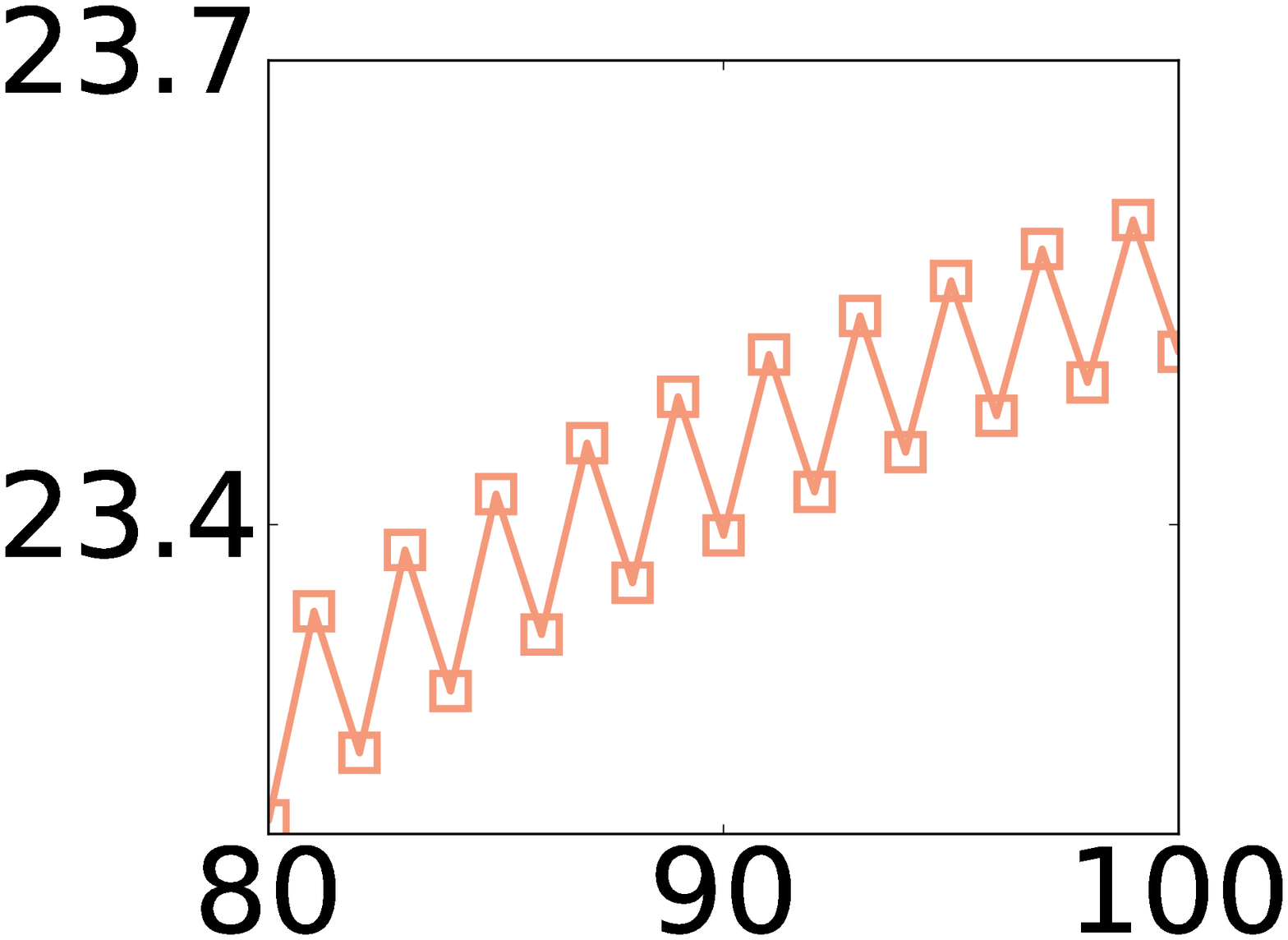}};
        \end{scope}
    \end{tikzpicture}
\label{WS_MSD_N_21}
}
\subfloat[]{\label{WS_MSD_N_201}\includegraphics[width=0.5\textwidth]{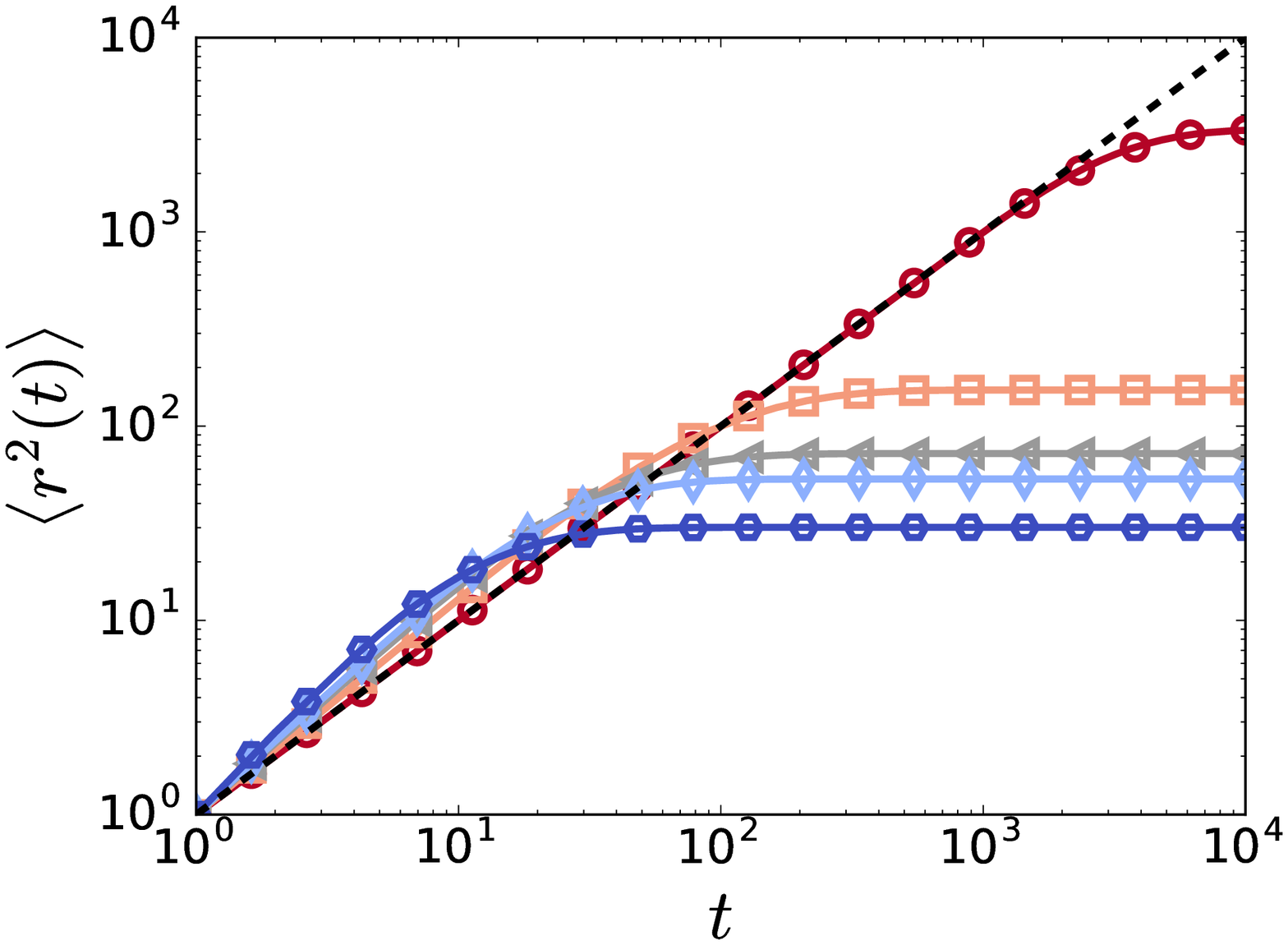}}
\caption{Time evolution of the MSD for a finite NW networks when $s=\infty$, and $p=0.0$ (red circles), $p=0.1$ (orange squares), $p=0.2$ (grey triangles), $p=0.3$ (cyan diamonds) and $p=0.5$ nodes (blue hexagons). In (a) $N=21$, and in (b) $N=201$. The inset of panel (a) illustrates the presence of small oscillations in the value of $\left \langle r^2(t) \right \rangle$ due to the ring topology. Such behavior is absent in the infinite linear chain. The black dashed line is a guide for the eye proportional to $t$.}
\label{WS_MSD_inf}
\end{figure}

\begin{figure}[h!]
\centering
\subfloat[]{\label{WS_der_N_21}\includegraphics[width=0.5\textwidth]{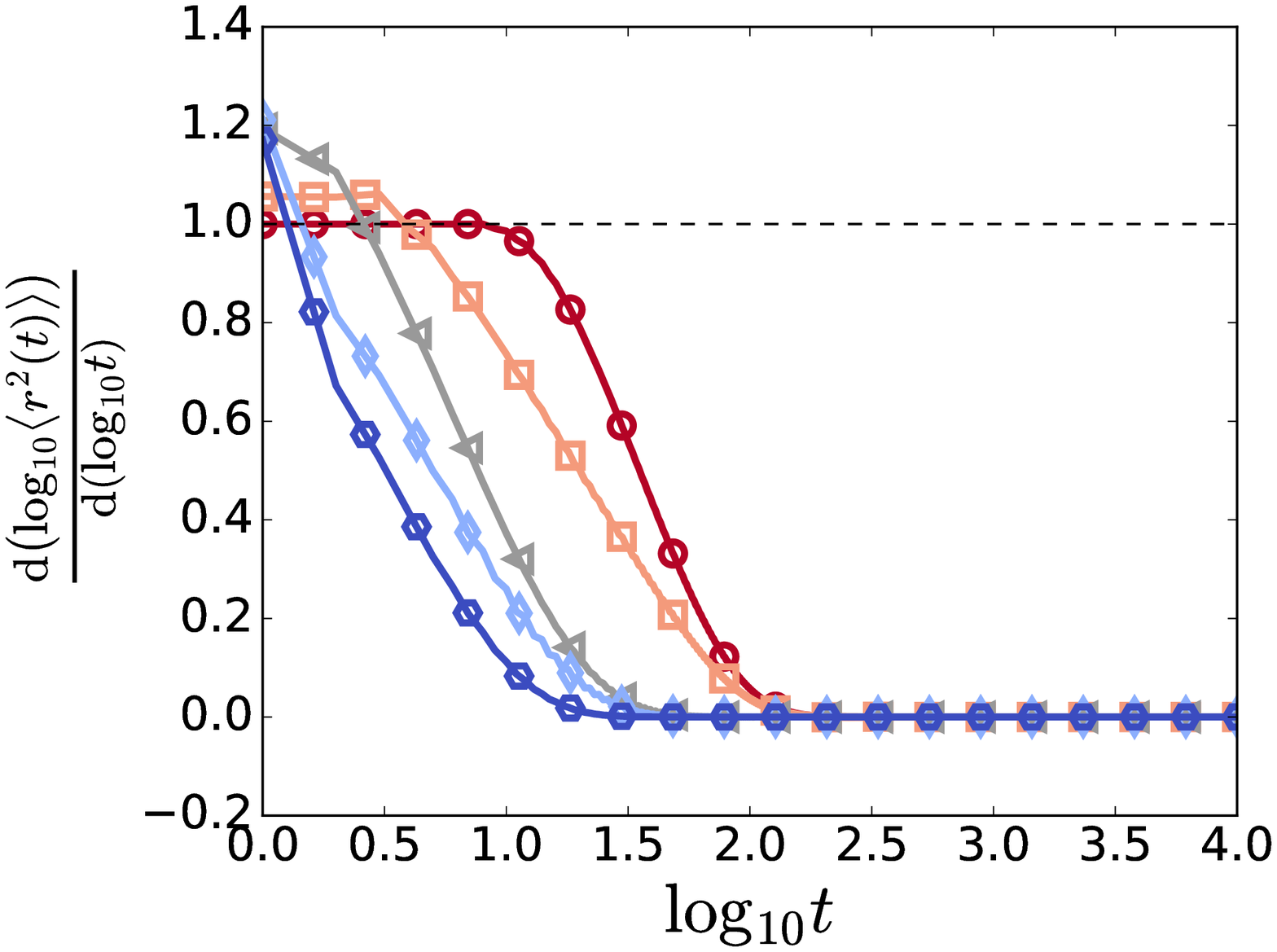}}
%\subfloat[]{\label{WS_der_N_101}\includegraphics[width=0.3\textwidth]{WS_der_N_101}}
\subfloat[]{\label{WS_der_N_201}\includegraphics[width=0.5\textwidth]{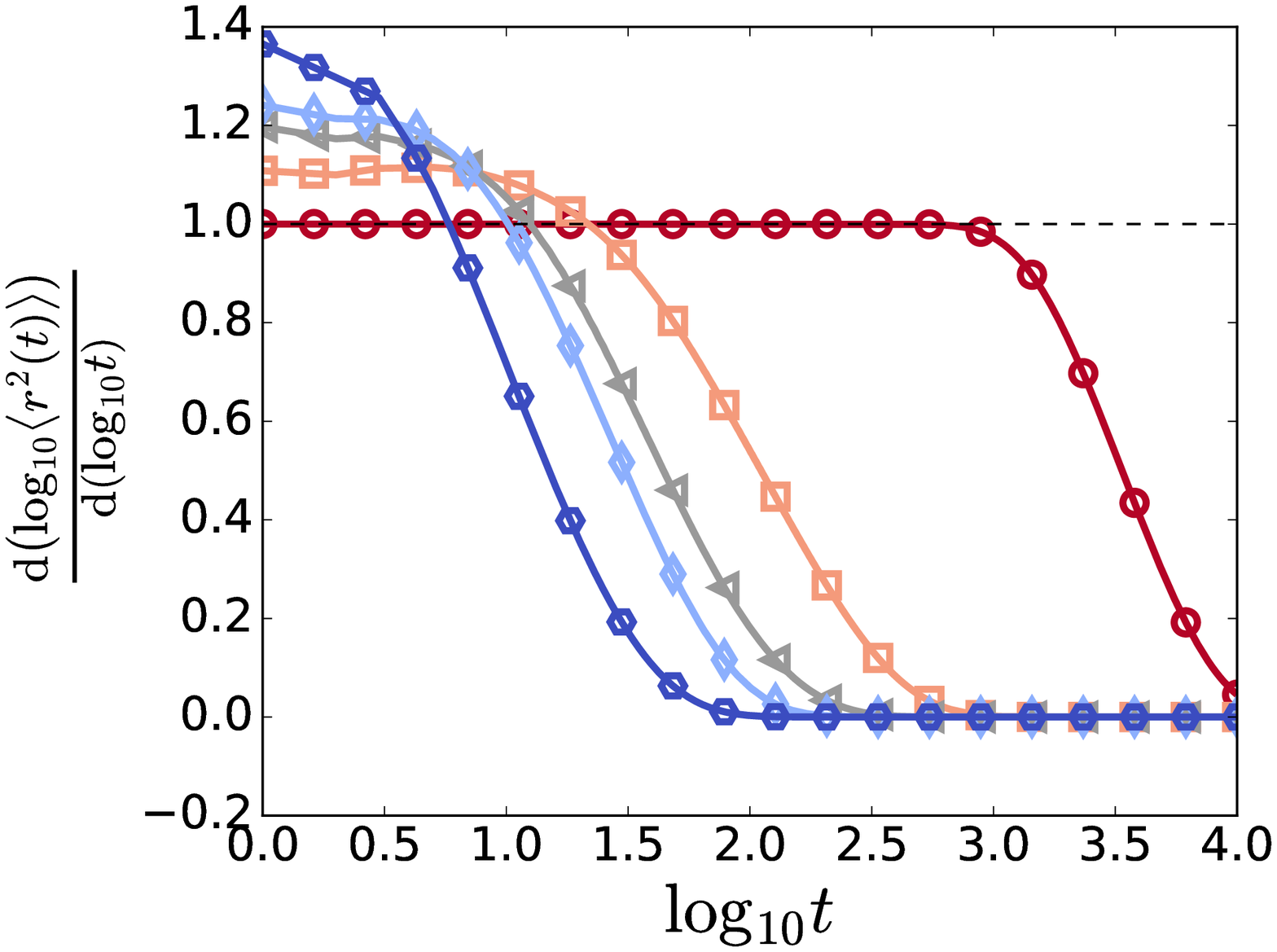}}
\caption{The results for the numerical derivative of $\log_{10} \left \langle r^2(t) \right \rangle$ with respect to $\log_{10}t$ obtained for the series in Fig.~\ref{WS_MSD_inf}(a) and Fig.~\ref{WS_MSD_inf}(b). The symbols are the same as in the referenced figures. The black dashed line is a guide for the eye to locate a normal diffusion. To avoid the presence of large oscillations due to the magnification of the oscillatory behavior shown in the inset of Fig.~\ref{WS_MSD_inf}(a), the derivatives were evaluated by taking the difference of the corresponding values at $t$ and $t+2$.}
\label{WS_der_inf}
\end{figure}

In Fig.~\ref{WS_MSD_inf}, we show some numerical estimations of Eq.~\ref{eq_MSD_multpl_new} on finite NW networks for various values of $p$, when $N=21$ and $N=201$, and when $s=\infty$. As can be observed, given a size $N$, the smaller the network diameter (i.e., the larger the value of $p$), the faster the saturation of MSD. Likewise, given a value of $p$, the smaller the $N$, the faster the saturation. On the other hand, when $p>0$, it is easy to see that the initial slope of $\left \langle r^2(t) \right \rangle$ is greater than 1, before saturation takes place. If $p$ is large and $N$ small, saturation may start already at time $t=2$ (see the results for $p=0.5$ in Fig.~\ref{WS_MSD_N_21}). In the case of $p=0$, we recover the features of cycle graphs (see Eqs. \ref{eq_MSD_s_cons} and \ref{approx_s_cons}). 
To obtain a more precise evaluation of the value of $\gamma$, in Fig.~\ref{WS_der_inf}, we present the results for the numerical derivative of $\log_{10} \left \langle r^2(t) \right \rangle$ with respect to $\log_{10}t$ of the series in Fig.~\ref{WS_MSD_inf}. Once the used Markovian formalism only provides MSD values for discrete time-steps, the derivatives were obtained by taking the quotient between the differences of the corresponding quantities taken at neighboring integer values of $t$.% \textcolor{red}{(see Eq.~\ref{Eq_der_NEW})}. %(see Appendix D for further informations about the methodology to estimate the log-derivatives).

%\textcolor{red}{
%\begin{eqnarray}
%\frac{\mathrm{d}\log_{10}(\left \langle r^2(t) \right \rangle) }{\mathrm{d} \log_{10}(t)}\approx \frac{\log_{10}(\left \langle r^2(t+2) \right \rangle)-\log_{10}(\left \langle r^2(t) \right \rangle)}{\log_{10}(t+2)-\log_{10}(t)}
%\label{Eq_der_NEW}
%\end{eqnarray}
%}

%%%%%%%%%%%%%%%%%%%%%%%%%%%%%%%%%%%%
%%%%%%%%%%%%%%%%%%%%%%%%%%%%%%%%%%%%

%%%%%%%%%%%%%%%%%%%%%%%%%%%%%%%%%%%%
%%%%%%%%%%%%%%%%%%%%%%%%%%%%%%%%%%%%%%%%%%%%%%%%%%%%%%

\begin{figure}[h!]
\centering
\subfloat[]{\label{WS_MSD_p_0p1}\includegraphics[width=0.5\textwidth]{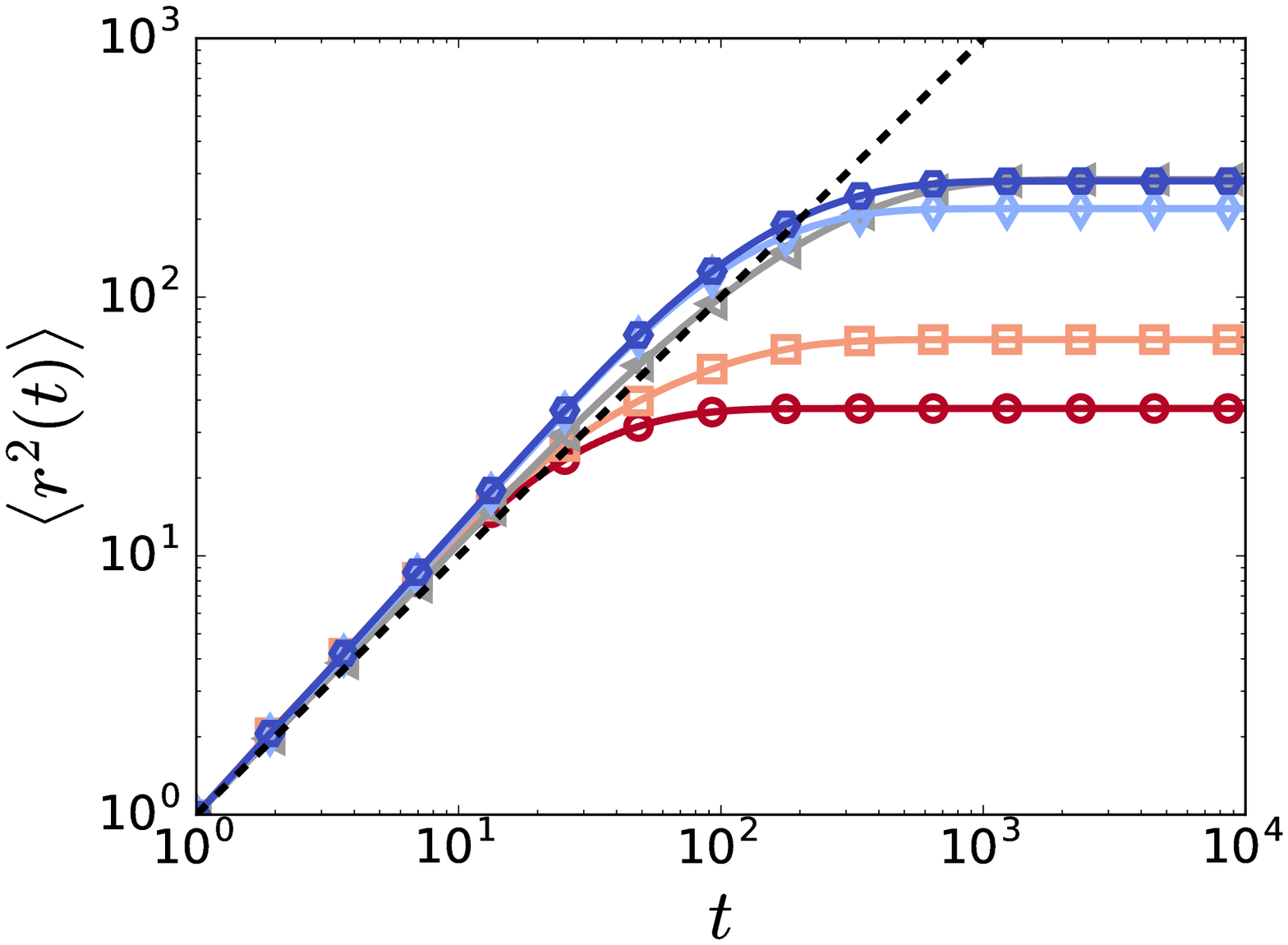}}
%\subfloat[]{\label{WS_der_N_101}\includegraphics[width=0.3\textwidth]{WS_der_N_101}}
\subfloat[]{\label{WS_der_p_0p1}\includegraphics[width=0.5\textwidth]{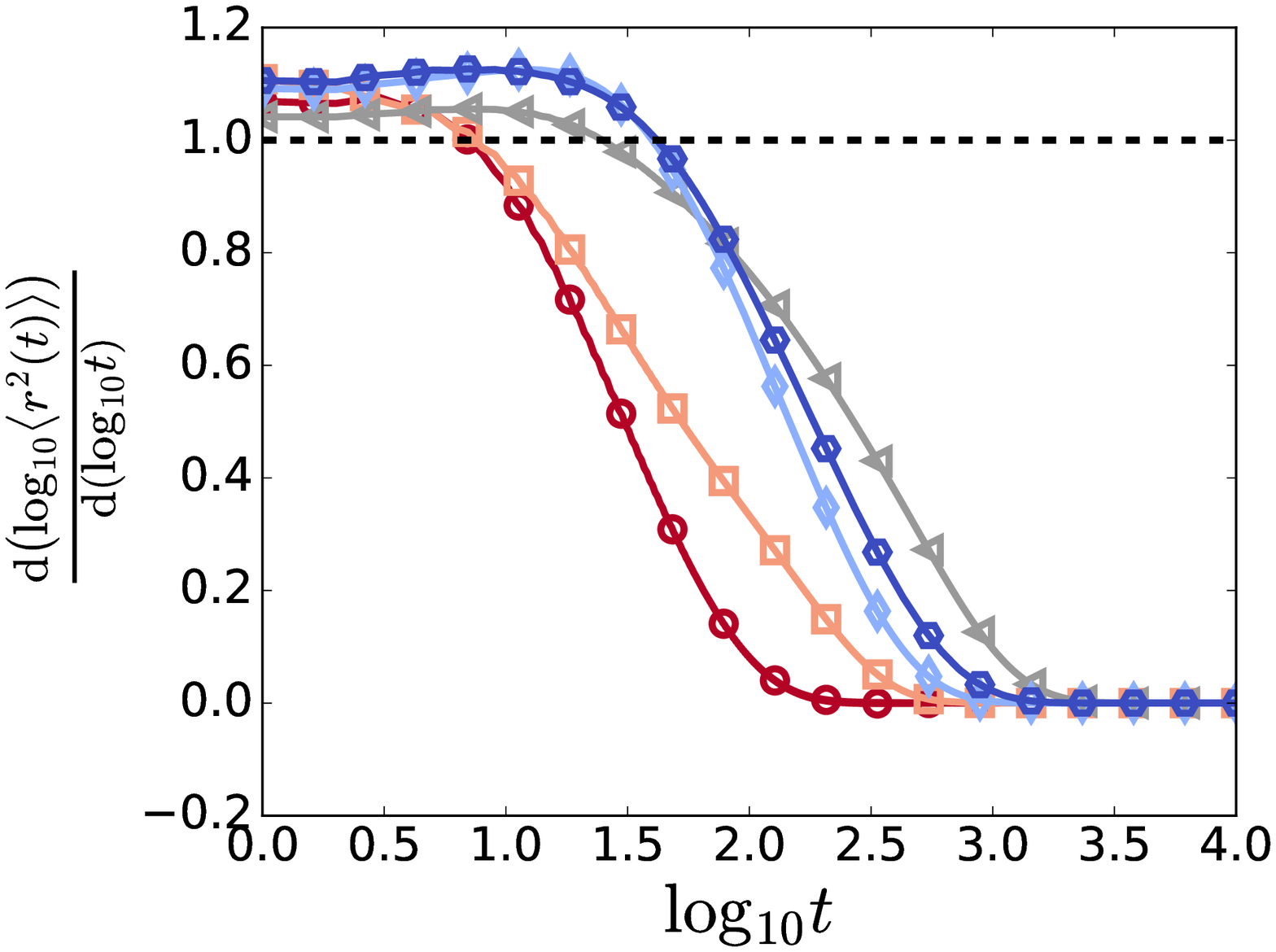}}
\caption{(a) Time evolution of the MSD for finite NW networks, when $s=\infty$ (i.e., there are no LRIs), $p=0.1$, and $N=33$ (red circles), $N=65$ (orange squares), $N=129$ (grey triangles), $N=257$ (cyan diamonds) and $N=513$ nodes (blue hexagons). (b) Numerical derivative of $\log_{10} \left \langle r^2(t) \right \rangle$ with respect to $\log_{10}t$ obtained for the series in Fig.~\ref{WS_inf_p_0p1}(a). The symbols are the same as in the referenced figure. The purple line indicates the results obtained by Eq.~\ref{eq_MSD_s_cons} for a cycle graph  ($p=0$), when $N=1025$ and $s=\infty$, which correspond to normal diffusion.}
\label{WS_inf_p_0p1}
\end{figure}

The results indicate that the conditions for superdiffusive behavior (i.e. $\left \langle r^2(t) \right \rangle \sim t^\gamma$ with $\gamma>1$) before saturation is satisfied in a variety of conditions. For the very small network with just 21 nodes, it is possible to notice superdiffusion during 3 time-steps when p = 0.1. This is a remarkable observation, which contrasts with the normal diffusion observed on cycle graphs before saturation, even when strong LRIs are introduced and N is very large. For larger values of $p$, the situation becomes unclear because of the very small network diameter. In Fig.~\subref*{WS_der_N_201} we see that, for N as large 201, $\gamma>$1 for larger time intervals and larger values of $p$. In particular, for $p=0.2$, we observe $\gamma>1$ until $t=4$. By way of contrast, previous results for supperdiffusion in finite systems caused by distant jumps were obtained for much larger systems ($N > 10^4$ nodes, see \cite{almaas03,gallos04}).  To support the robustness of the above picture, in Fig.~\ref{WS_inf_p_0p1} we show results for $p=0.1$ and increasing values of $N \in[33,513]$. As can be seen, the condition $\gamma>1$ is consistently satisfied for larger and larger values of $t$ as $N$ is increased.

%As can be seen in Fig.~\ref{WS_der_inf}, the variation of the parameter $p$ reveals that a superdiffusive-like behavior emerges (i.e. $\left \langle r^2(t) \right \rangle \sim t^\gamma$ with $\gamma>1$), before saturation, when the amount of shortcuts is very small ($p=0.1$). This is a remarkable observation, which contrasts with the normal diffusion observed on cycle graphs before saturation, even when LRIs are introduced. On the other hand, the size of discrete superdiffusive systems (where discrete jumps or \enquote{hops} take place) is very large ($N \gtrsim 10^4$ nodes, see \cite{almaas03,gallos04}). However, in Fig.~\subref*{WS_der_N_21}, it is possible to see superdiffusion on a network with just 21 nodes, during 4 time-steps. In Fig.~\ref{WS_inf_p_0p1}, we show more examples of the time evolution of the MSD for various finite NW networks, when $p=0.1$.  As can be seen, when $p=0.1$ and there are no LRIs, $1<\gamma \lesssim 1.1 $, and the larger the diameter of the NW network (i.e., the larger the size of the system, $N$), the larger the duration of this superdiffusive regime.

%%%%%%%%%%%%%%%%%%%%%%%%%%%%%%%%%%%%
%%%%%%%%%%%%%%%%%%%%%%%%%%%%%%%%%%%%%%%%%%%%%%%%%%%%%%

\begin{figure}[h!]
\centering
\subfloat[]{\label{WS_MSD_s_cons_varios_s}\includegraphics[width=0.5\textwidth]{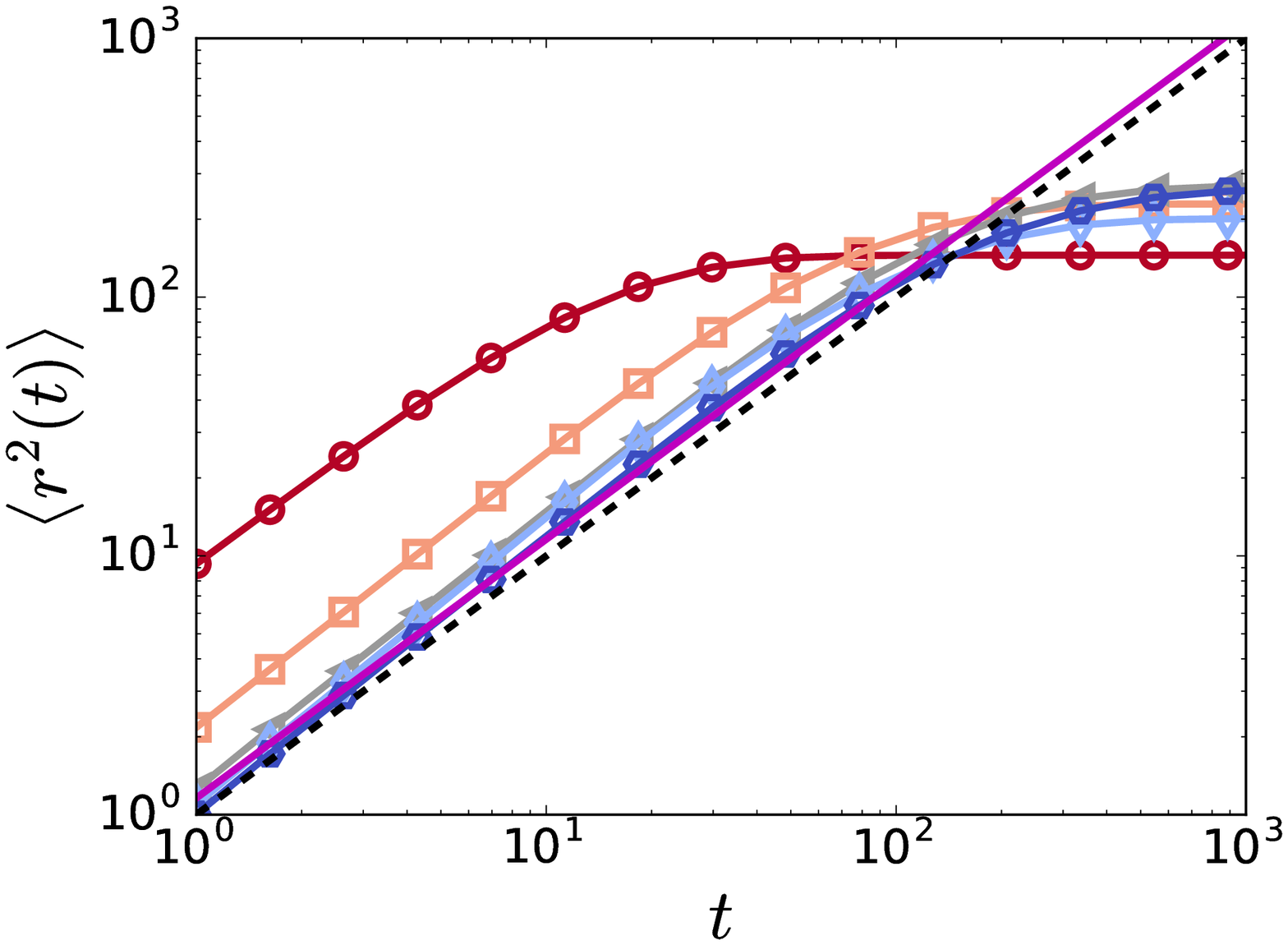}}
\subfloat[]{\label{WS_Der_s_cons_varios_s}\includegraphics[width=0.5\textwidth]{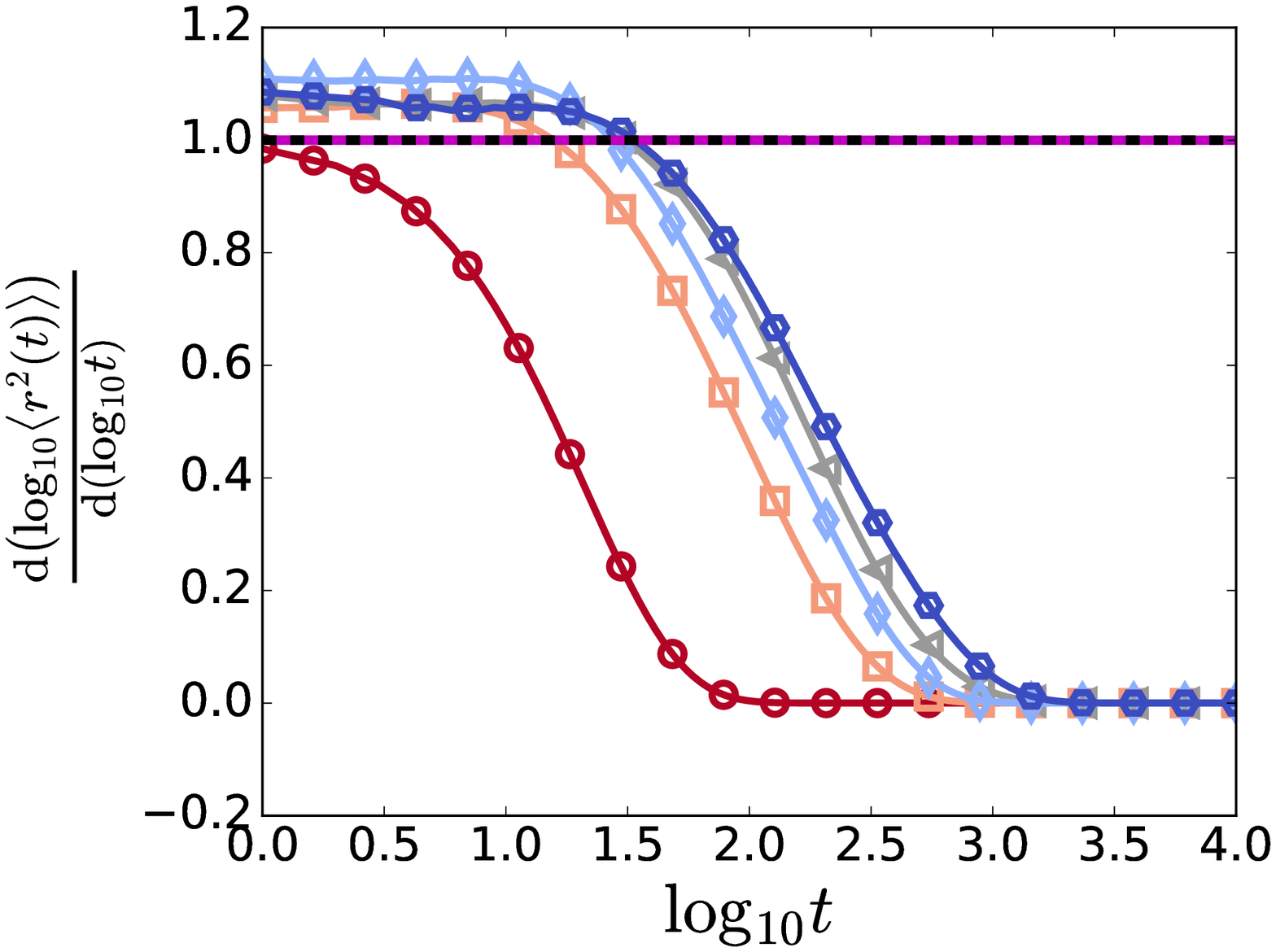}}
%\subfloat[]{
 %   \begin{tikzpicture}
 %       \node[anchor=south west,inner sep=0] (image) at (0,0) {\includegraphics[width=0.5\textwidth]{WS_Der_s_cons_varios_s}};
%        \begin{scope}[x={(image.south east)},y={(image.north west)}]
   %         %\draw[help lines,xstep=.1,ystep=.1] (0,0) grid (1,1);
%            %\foreach \x in {0,1,...,9} { \node [anchor=north] at (\x/10,0) {0.\x}; }
%            %\foreach \y in {0,1,...,9} { \node [anchor=east] at (0,\y/10) {0.\y}; }
  %          \node[anchor=south west,inner sep=0] (image) at (0.23,0.195) {\includegraphics[width=0.155\textwidth]{inset_der_257_p01}};
%        \end{scope}
 %   \end{tikzpicture}
%\label{WS_Der_s_cons_varios_s}
%}
\caption{(a) Time evolution of the MSD for finite NW networks with LRIs, when $p=0.1$, $N=257$, and $s=3$ (red circles), $s=4$ (orange squares), $s=5$ (grey triangles), $s=6$ (cyan diamonds) and $s=10$ (blue hexagons). (b) Numerical derivative of $\log_{10} \left \langle r^2(t) \right \rangle$ with respect to $\log_{10}t$ obtained for the series in Fig.~\ref{fig_WS_MSD_s_cons_varios_s}(a). The symbols are the same as in the referenced figure. The purple line indicates the results obtained by Eq.~\ref{eq_MSD_s_cons} for a cycle graph  ($p=0$), when $N=1025$ and $s=5$. The black dashed lines are a guides for the eye to locate a normal diffusion.}
\label{fig_WS_MSD_s_cons_varios_s}
\end{figure}

\begin{figure}[h!]
\centering
\subfloat[]{\label{WS_MSD_s_cons_varios_s_p05}\includegraphics[width=0.5\textwidth]{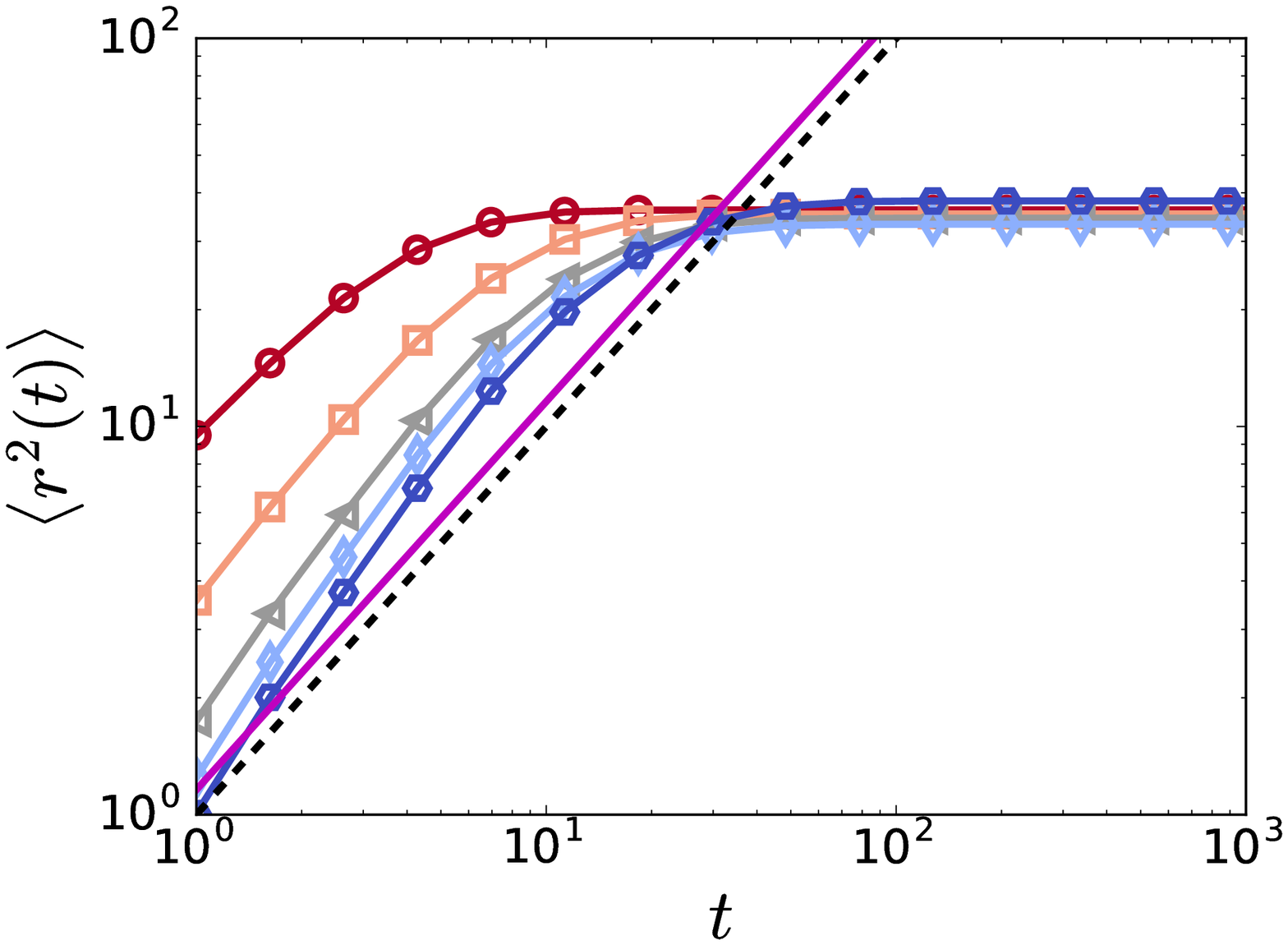}}
%\subfloat[]{\label{WS_der_N_101}\includegraphics[width=0.3\textwidth]{WS_der_N_101}}
\subfloat[]{\label{WS_der_s_cons_varios_s_p05}\includegraphics[width=0.5\textwidth]{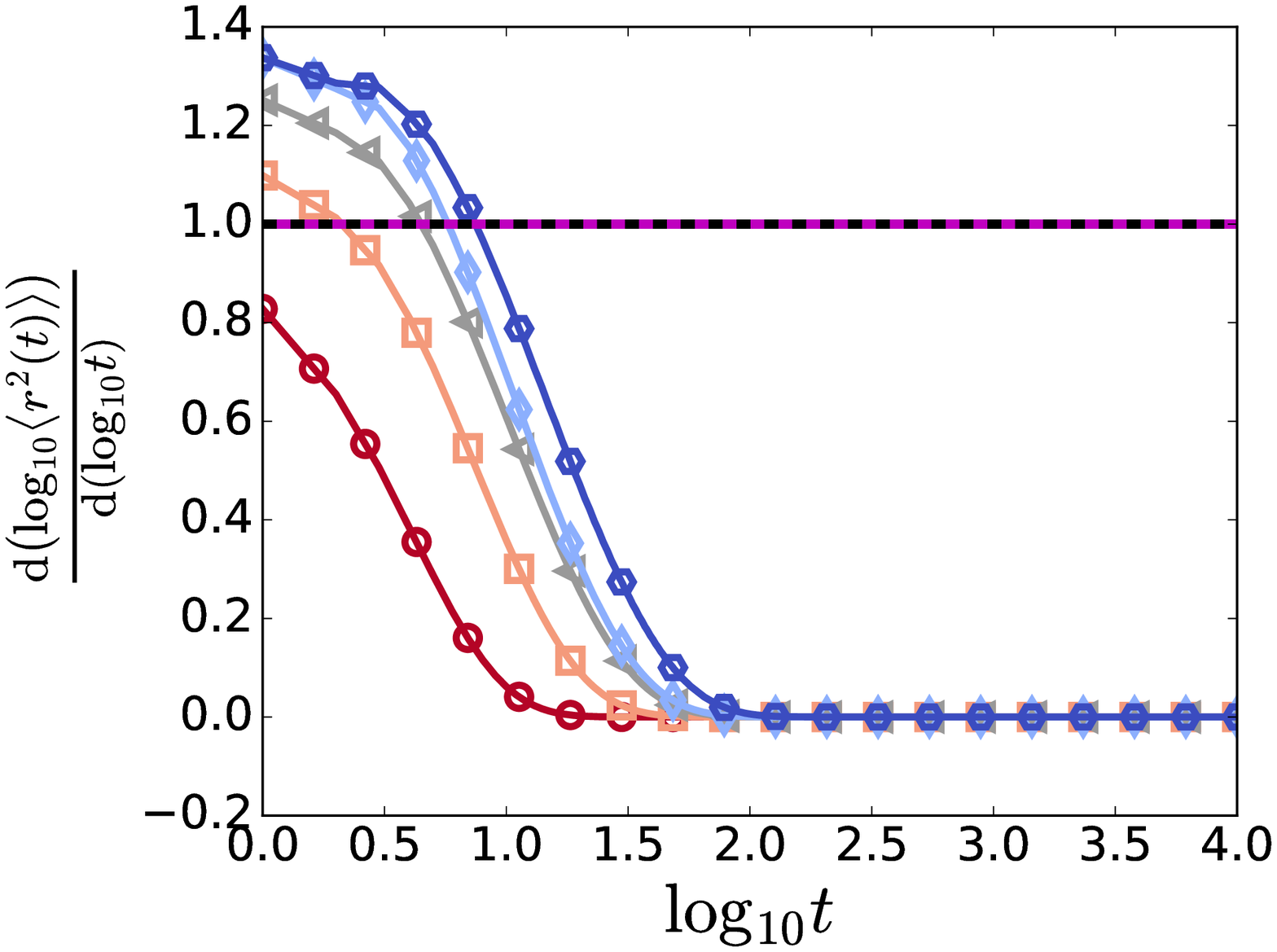}}
\caption{(a) Time evolution of the MSD for finite NW networks with LRIs, when $p=0.5$, $N=257$, and $s=3$ (red circles), $s=4$ (orange squares), $s=5$ (grey triangles), $s=6$ (cyan diamonds) and $s=10$ (blue hexagons). (b) Numerical derivative of $\log_{10} \left \langle r^2(t) \right \rangle$ with respect to $\log_{10}t$ obtained for the series in Fig.~\ref{Fig_WS_MSD_s_cons_varios_s}(a). The symbols are the same as in the referenced figure. The purple line indicates the results obtained by Eq.~\ref{eq_MSD_s_cons} for a cycle graph  ($p=0$), when $N=1025$ and $s=5$. The black dashed lines are a guides for the eye to locate a normal diffusion.}
\label{Fig_WS_MSD_s_cons_varios_s}
\end{figure}

The inclusion of LRIs in NW networks impacts the described superdiffusive behavior. In Figs. \ref{fig_WS_MSD_s_cons_varios_s} and \ref{Fig_WS_MSD_s_cons_varios_s} we show the results obtained for a fixed system size ($N = 257$ nodes) and various values of $s<\infty$ for, respectively, $p = 0.1$ and $p = 0.5$. As can be observed from a comparison of the figures, the role played by $s$ and $p$ become similar, in the sense that increasing the strength of the LRIs has the same effect as adding extra connections: the saturation regime is reached at an increasing faster pace. We notice that very intense LRIs, corresponding e.g. to $s =3$, prevent the evaluation of $\gamma >1$ for $p=0.1$, although a short superdiffusive interval occurs when $p$ is reduced to 0.05. The optimal condition for superdiffusion is observed for small $p$ (but necessarily $>0$) and large values of $s$.

%We study numerically the diffusive behavior of the finite NW networks with LRIs. In Figs. \ref{fig_WS_MSD_s_cons_varios_s} and \ref{Fig_WS_MSD_s_cons_varios_s}, we show the results obtained for a given system size ($N=257$ nodes) and various values of the Mellin parameter ($s \ll \infty$, i.e., there are LRIs), when $p=0.1$ (i.e., there are very few shortcuts) and $p=0.5$  (i.e., there are many shortcuts), respectively. As can be observed in both figures, the smaller the value of $s$, the faster the saturation. Indeed, very intense LRIs (i.e., $s < 4$) prevent the appearance of superdiffusion, even in the case of $p=0.1$ (see the inset in Fig.~\subref*{WS_Der_s_cons_varios_s}). Therefore, to observe a lasting superdiffusive behavior on finite NW networks, first of all, the system should exhibit that feature without cosntant LRIs (i.e., $p\approx 0.1$ and $s=\infty$) and, then, the LRIs should not be very intense.

%%%%%%%%%%%%%%%%%%%%%%%%%%%%%%%%%%%%
%%%%%%%%%%%%%%%%%%%%%%%%%%%%%%%%%%%%%%%%%%%%%%%%%%%%%%

%%%%%%%%%%%%%%%%%%%%%%%%%%%%%%%%%%%%
%%%%%%%%%%%%%%%%%%%%%%%%%%%%%%%%%%%%%%%%%%%%%%%%%%%%%%

As can be observed in Figs. \ref{WS_inf_p_0p1}-\ref{Fig_WS_MSD_s_cons_varios_s}, our numerical results for NW networks with very few shortcuts (i.e., $p=0.1$) clearly indicate that the exponent $\gamma$ and the power-law regime duration, $t_f$, depend on $N$ and $s$, respectively. In order to study the conditions that allow the emergence of a lasting superdiffusive behavior on these systems, finally, we explore the dependence of $\gamma$ and of $t_f$, on the previous parameters.  To do so, we first consider that, before saturation takes place, the exponent $\gamma$ is approximately equal to $\gamma_0$, the estimated value of the derivative of $\log_{10} \left \langle r^2(t) \right \rangle$ with respect to $\log_{10}t$ at $t=1$. On the other hand, we consider that $t_f$ is proportional to $t_{\mathrm{lim}}$, the first value of $t$ where the derivative of $\log_{10} \left \langle r^2(t) \right \rangle$ with respect to $\log_{10}t$ is less than or equal to $0.96\gamma_0$, a value that still satisfies the condition $\gamma>1$ if $\gamma_0>1.042$. Given these criteria, we study the dependence of $\gamma_0$ and $t_{\mathrm{lim}}$ on $N$ and $s$.

\begin{figure}[h!]
\centering
\subfloat[]{\label{NW_gamma_0}\includegraphics[width=0.5\textwidth]{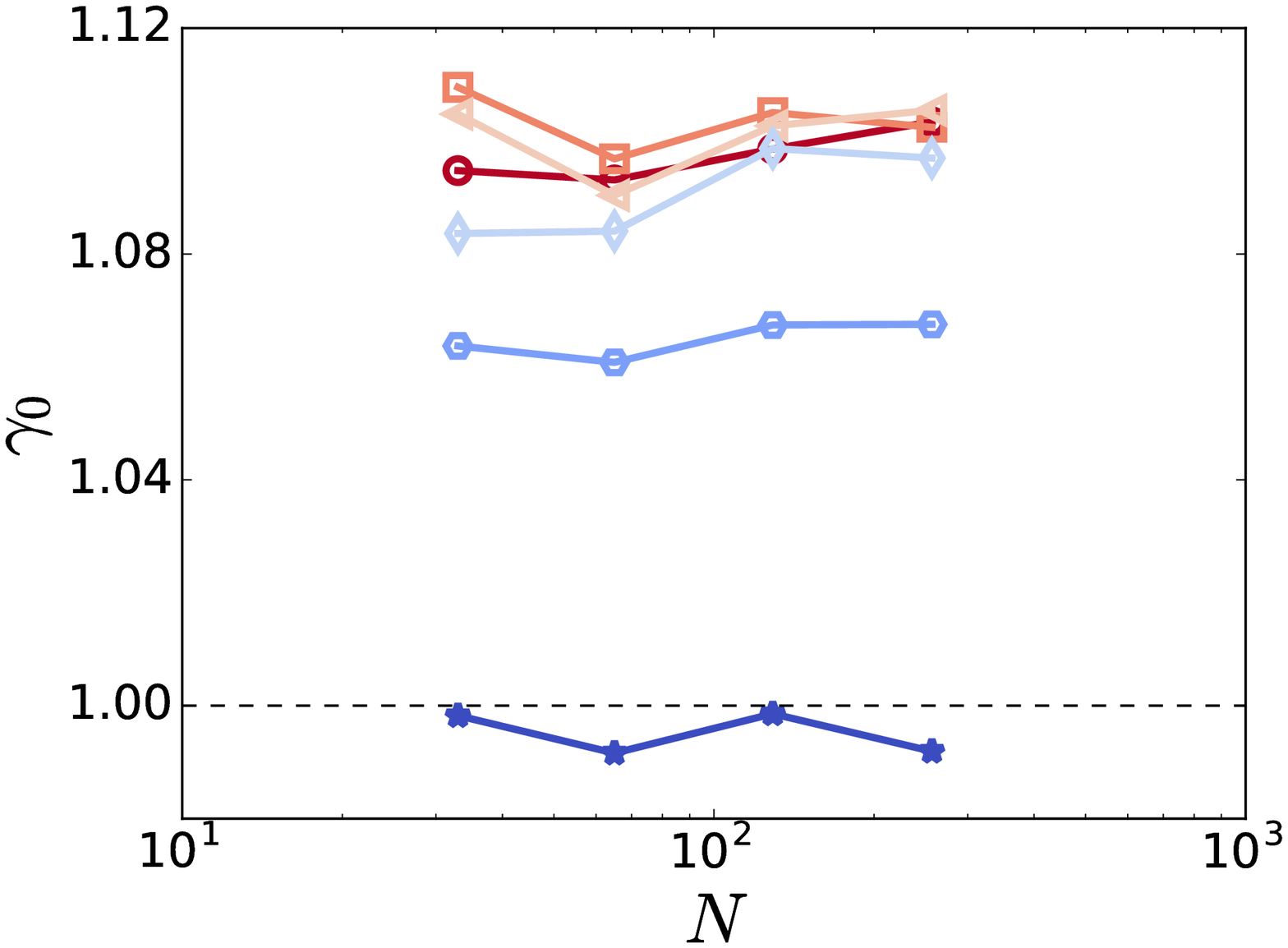}}
\subfloat[]{\label{r_ini_p01}\includegraphics[width=0.5\textwidth]{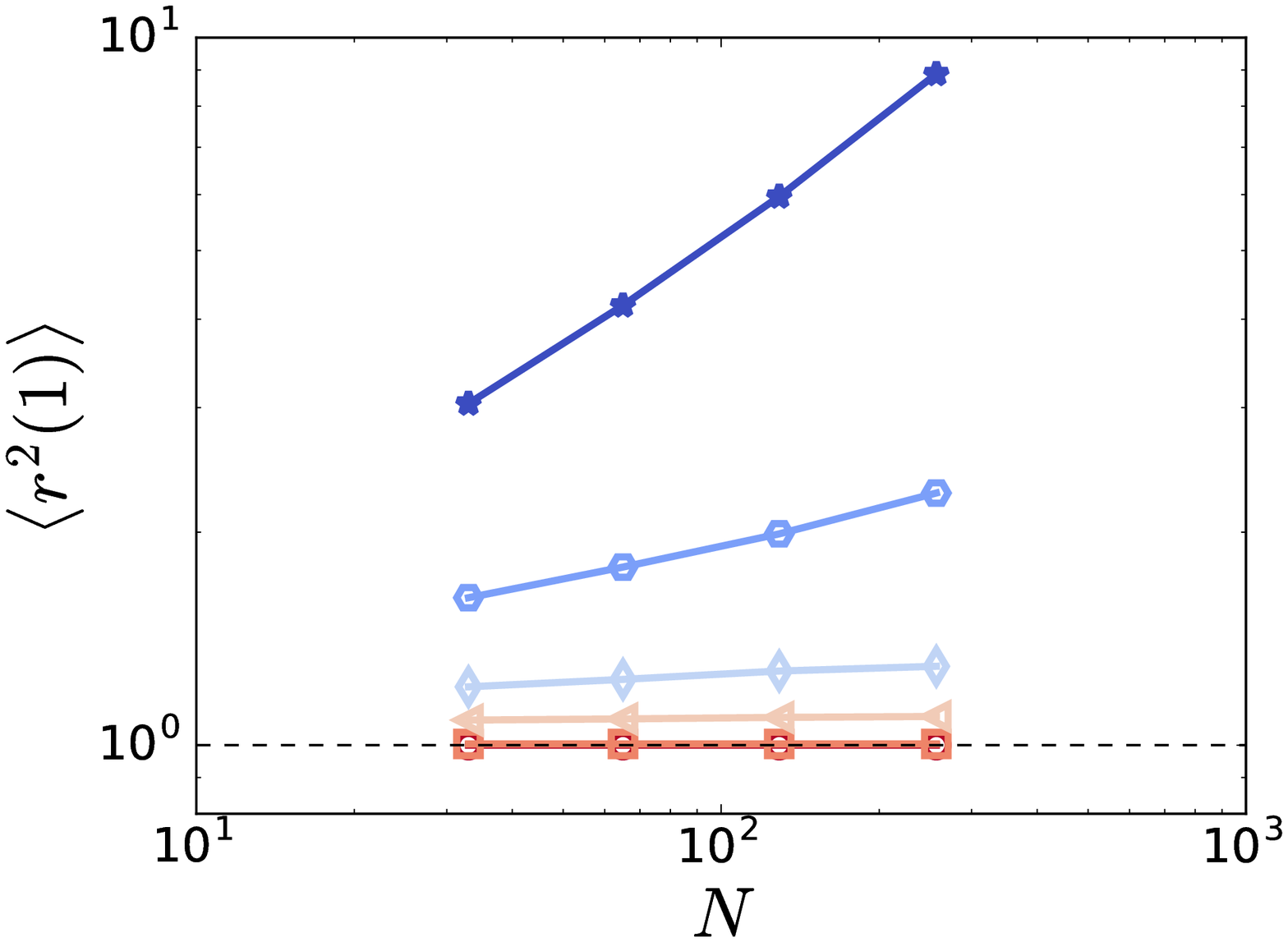}}

\medskip

\subfloat[]{\label{ratio_p01}\includegraphics[width=0.5\textwidth]{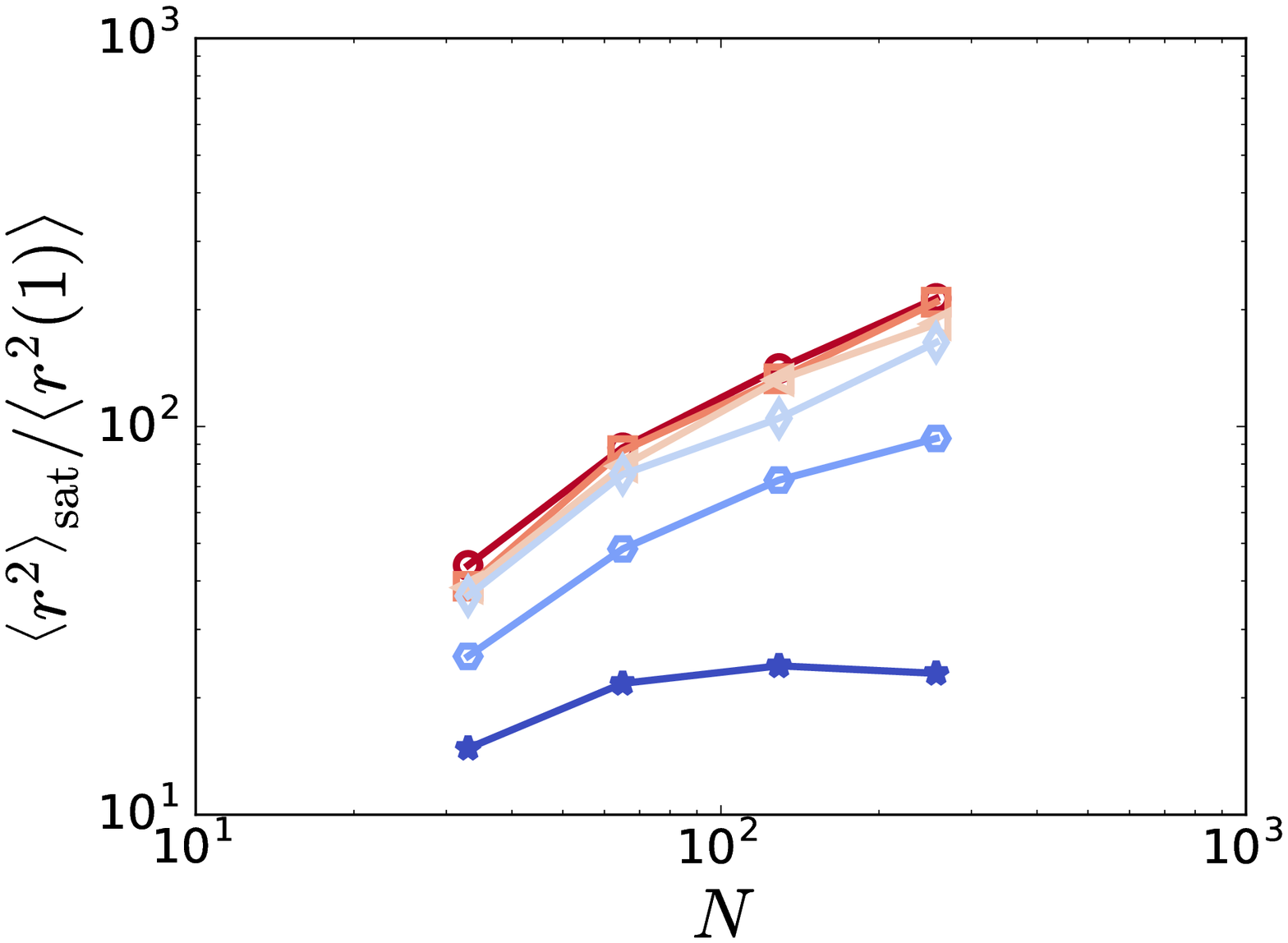}}
\subfloat[]{\label{ini_N_257_varios_p}\includegraphics[width=0.5\textwidth]{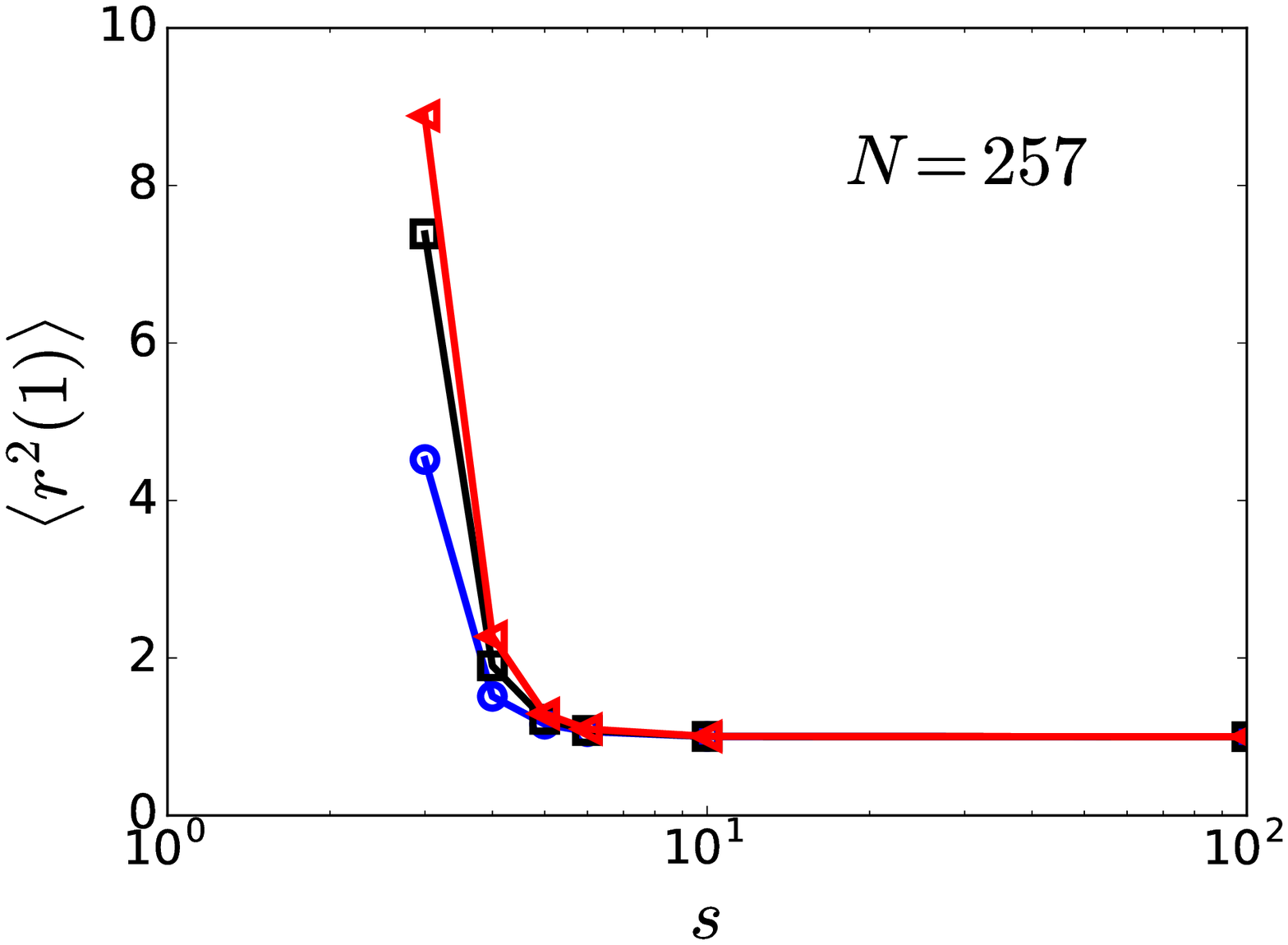}}
\caption{Results (averaged over 100 realizations) for $\gamma_0$ (a), $\left \langle r^2(1) \right \rangle$ (b), and $\left \langle r^2 \right \rangle_\mathrm{sat}/\left \langle r^2(1) \right \rangle$ (c) for NW networks with LRIs, when $p=0.1$ and $N=33$, 65, 129, and 257: $s=\infty$ (circles), 10 (squares), 6 (triangles), 5 (diamonds), 4 (hexagons) and 3 (asterisks). In (a) and (b), the black dashed line represents the corresponding values of normal diffusion. (d) Results for $\left \langle r^2(1) \right \rangle$ (averaged over 100 realizations) for NW networks with $N=257$, when $p=0.00$ (circles), 0.05 (squares), and 0.10 (triangles).}
\label{results_NW_p01}
\end{figure}

In Fig~\subref*{NW_gamma_0}, we show the results for $\gamma_0$ (averaged over 100 realizations). As can be observed, for $s\geq5$ we obtain quite similar values $\gamma_0\approx 1.1$. In the case of $s\lesssim 5$, for a given size $N$, the smaller the value of $s$, the smaller $\gamma_0$. %On the other hand, when $0<s\lesssim 5$, the curves of constant $s$ are concave and have their maximum at $N=65$ or 129.

Additionally, it is possible to see in Fig~\subref*{NW_gamma_0} that, when the LRIs become very intense (i.e., $s=3$), $\gamma_0\lesssim 1$. In this case, superdiffusion can not be identified by the adopted criteria $\gamma_0>1.042$. As can be observed in Fig.~\subref*{r_ini_p01}, when the strength of LRIs increases and $s$ approaches the value 3, $\left \langle r^2(1) \right \rangle$ also increases significantly, becoming very close to $\left \langle r^2 \right \rangle_\mathrm{sat}$ (see in Fig.~\subref*{ratio_p01}). We do not observe this behaviour in cycle graphs ($p=0.0$) once, despite a large increase in the value of $\left \langle r^2(1) \right \rangle$, it still stays sufficiently away from $\left \langle r^2 \right \rangle_\mathrm{sat}$ (see in Fig.~\subref*{ini_N_257_varios_p}).

%%%%%%%%%%%%%%%%%%%%%%%%%%%%%%%%%%%%

\begin{figure}[h!]
\centering
\includegraphics[width=0.5\textwidth]{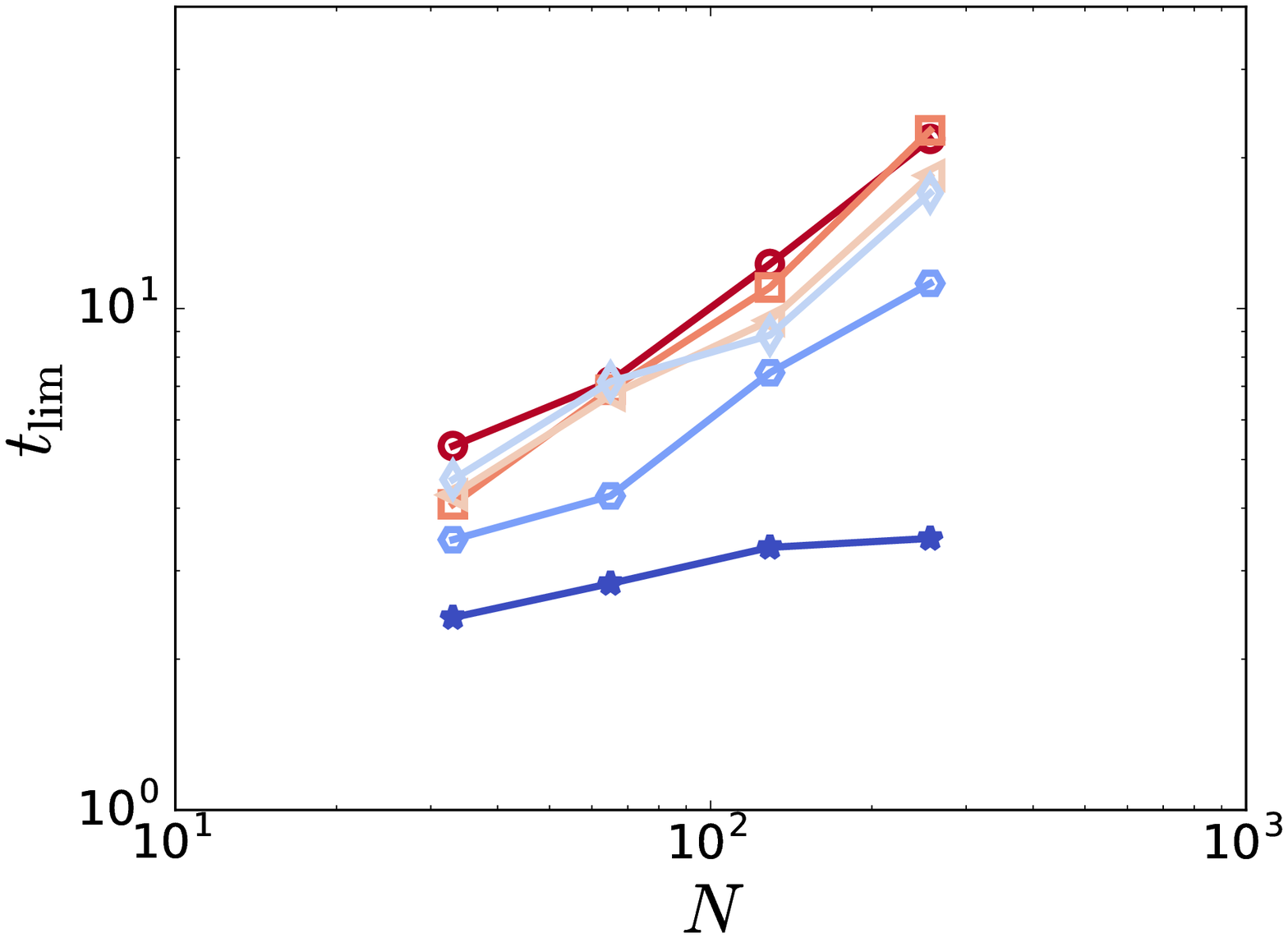}
\caption{Results (averaged over 100 realizations) for $t_{\mathrm{lim}}$ on NW networks with LRIs, when $p=0.1$ and $N=33$, 65, 129, and 257: $s=\infty$ (circles), 10 (squares), 6 (triangles), 5 (diamonds), 4 (hexagons) and 3 (asterisks).}
\label{lNW_t_lim}
\end{figure}

Finally, in Fig.~\ref{lNW_t_lim} we exhibit the results for $t_{\mathrm{lim}}$ as a function of $N$ (averaged over 100 realizations). As can be seen, for any value of $s$, the larger the system size, the larger the value of $t_{\mathrm{lim}}$. On the other hand, it is possible to observe that the results for $s\geq5$ become quite similar. Also, in accordance with previous discussions, very intense LRIs (i.e., $s < 4$) reduce the value of $t_{\mathrm{lim}}$ for any given $N$ and, consequently, prevent the appearance of superdiffusion.

%%%%%%%%%%%%%%%%%%%%%%%%%%%%%%%%%%%%
%%%%%%%%%%%%%%%%%%%%%%%%%%%%%%%%%%%%

%Finally, comparing the curves in Fig.~\ref{WS_Der_s_cons_varios_s} with those in Figs. \ref{WS_der_p_0p1} and \ref{WS_Der_s_cons_varios_s}, it is possible to observe that LRIs reduce the fluctuations of the log-derivatives that take place during its saturation process.

%%%%%%%%%%%%%%%%%%%%%%%%%%%%%%%%%%%%%%%%%%%%%%%%%%%%%%%%%%%%%%%%%%%%%%%%%%%%%%%%%%%%%%%%%%%%%%%%%%%%%%%%%%%%%%%%%%%%%%%%%%%%%%%%%%%%%%%%%%%%%%%%%%%%%%%%%%%%%%%%%%%%%%%%

\section{Conclusions}
\label{Con}

In this work, we have studied the time dependence of MSD for diffusion processes in complex networks, by comparing the differences caused by the inclusion of new connections between the nodes with those produced by the presence of LRIs. To do so, we have extended the formalism of MSD estimation to finite networks with $d-$path transformations. As it has been shown, given a $d-$path transformation $\tau$, the result depends on several factors: the discrete time-step $t$, the weight of the $d-$path transformation, and the topology of the network. For the sake of simplicity, in this work we only have considered the case of $\tau=\mathrm{Mellin}$.

To have a more clear picture of the effects of the weight $s$ of the Mellin transformation and of the topology, our approach has been specifically adapted to finite cycle graphs and NW networks. On the one hand, cycle graphs avoid the typical fast saturation of the MSD, and, additionally, they also have allowed us to derive analytical expressions for several dynamical features of the process. Then, we have used the NW model to add shortcuts to ordered cycle graphs and, thus, create small-world networks.

Our findings show that a super-diffusive-like behavior can emerge when shortcuts are added to finite cycle graphs, whether the systems have LRIs or not. We have verified that this feature is due to the introduction of the new connections. In the case of finite cycle graphs, our numerical and analytical results indicate that the MSD on these systems exhibits a normal diffusion ($\gamma=1$), before saturation takes place. In fact, an universal curve for the MSD evolution has been obtained. This shows that the presence of weaker non-local interactions among all pairs of nodes that are not directly connected to each other enhances the diffusion velocity but does not change the linear relation between MSD and $t$.

In the case of NW networks without LRIs (i.e, $s = \infty$), the variation of the parameter $p$ reveals that a lasting superdiffusive-like behavior emerges (i.e. $\left \langle r^2(t) \right \rangle \approx t^\gamma$ with $\gamma>1$), before saturation appears, when the amount of shortcuts is very small ($p \approx 0.1$). Our numerical results for $p=0.1$ indicate that $1<\gamma \lesssim 1.1 $, and the larger the diameter of the NW network (i.e., the larger the size of the system, $N$), the larger the duration of this superdiffusive regime.

On the other hand, when significant LRIs are considered (i.e, $0 < s \lesssim 10$), the smaller the value of $s$, the faster the saturation. Indeed, very intense LRIs (i.e., $s < 4$) prevent the appearance of superdiffusion, even in the case of $p=0.1$. Therefore, to observe a lasting superdiffusive behavior on finite NW networks, first of all, the system should exhibit that feature without cosntant LRIs (i.e., $p\approx 0.1$ and $s=\infty$) and, then, the LRIs should not be very intense.

Preliminary investigations of the same system defined on medium-sized (connected) Watts-Strogatz networks and balanced tree graphs have led to qualitatively similar results to those obtained for the cycle graphs. Of course these systems are hardly amenable to exact analytical approaches, so that a thorough investigation must heavily rely on very large size samples. Thus, given the wide applicability of the $d-$path transformations to networked systems, this work constitutes the first step toward a better understanding the effect of non-local connections on diffusive processes on top of complex networks.

%%%%%%%%%%%%%%%%%%%%%%%%%%%%%%%%%%%%%%%%%%%%%%%%%%%%%%%%%%%%%%%%%%%%%%%%%%%%%%%%%%%%%%%%%%%%%%%%%%%%%%%%%%%%%%%%%%%%
%%%%%%%%%%%%%%%%%%%%%%%%%%%%%%%%%%%%%%%%%%%%%%%%%%%%%%%%%%

\begin{acknowledgments}

We gratefully acknowledge Ernesto Estrada, Evaldo Curado and Fernando Nobre for fruitful discussions.
This work was supported by the project MTM2015-63914-P from the Ministry of Economy and Competitiveness of Spain and by the Brazilian agencies CNPq and CAPES. RFSA also acknowledges the support of the National Institute of Science and Technology for Complex Systems (INCT-SC Brazil).
\end{acknowledgments}

%%%%%%%%%%%%%%%%%%%%%%%%%%%%%%%%%%%%%%%%%%%%%%%%%%%%%%%%

\section*{APPENDIX A: Eigenvalue spectrum of a circulant matrix.}
\setcounter{subsection}{0}
\label{appA}

A $N\times N$ circulant matrix $\mathbf{C}$ takes the form

\begin{equation}
\mathbf{C}=\begin{pmatrix}
c_0 & c_{N-1} & \cdots & c_{2} & c_{1}\\
c_1 & c_{0} & c_{N-1} &  & \\
\vdots & c_{1} & c_{0} & \ddots & \vdots\\
c_{N-2} &  & \ddots & \ddots & c_{N-1} \\
c_{N-1} & c_{N-2}& \cdots & c_{1} & c_{0} \\
\end{pmatrix}.
\end{equation}

A circulant matrix $\mathbf{C}$ is fully specified by one vector, $\vec{c}$, which appears as the first column of $\mathbf{C}$. The remaining columns of $\mathbf{C}$ are each cyclic permutations of the vector $\vec{c}$ with offset equal to the column index. 

The normalized eigenvectors of a circulant matrix are given by:
\begin{equation}
\vec{v}_j=\frac{1}{\sqrt{N}}\begin{pmatrix}
1 & \omega_j  & \omega_j^{2} & \cdots & \omega_j^{N-1}
\end{pmatrix}^T,
\end{equation}

\noindent for $j=0,1,\cdots,N-1$, where 

\begin{equation}
\omega_j=\exp\left ( \mathfrak{i}\frac{2\pi  j}{N} \right ),
\end{equation}

\noindent are the $N-$th roots of unity and $\mathfrak{i}$ is the imaginary unit.

The corresponding eigenvalues are then given by:

\begin{equation}
\sigma_j^{\mathbf{C}}=c_0+\sum_{k=1}^{N-1}c_k\omega_j^{N-k},
\end{equation}

\noindent for $j=0,1,\cdots,N-1$.
%%%%%%%%%%%%%%%%%%%%%%%%%%%%%%%%%%%%%%%%%%%%%%%%
\section*{APPENDIX B: MSD on weighted networks}
\label{Analytic_cicles}

\subsection{The length of the shortest path between two nodes of a weighted graph}

Since the $d-$path transformation of $G=(V,E)$ defines a weighted topology (see Eq.~\ref{adj_transf_1}), we can extend the methodology to estimate MSD on simple graphs. To do so, we take into consideration the influence of the weights of the ties on the length of the shortest path between two nodes. Following \cite{opsahla10}, we assume that the weights $\hat{\mathbf{A}}^\tau \left (i,j \right )$ are operationalizations of tie strength between the nodes $i$ and $j$. Thus, the length of the shortest path between two nodes of a weighted graph $G$, $d_{i,j}^\alpha$, can be formalized as:

\begin{equation}
 d_{i,j}^\alpha=\min\left ( \frac{1}{\left ( \hat{\mathbf{A}}^\tau \left (i,h_1 \right ) \right )^\alpha}+\frac{1}{\left ( \hat{\mathbf{A}}^\tau \left (h_1,h_2 \right ) \right )^\alpha}+\cdots+\frac{1}{\left ( \hat{\mathbf{A}}^\tau \left (h_\ell,j \right ) \right )^\alpha} \right ),
\end{equation}

\noindent where $h_1$, $h_2$, $\cdots$, $h_\ell$ are intermediary nodes on a given path between node $i$ and $j$, $\alpha$ is a nonnegative tuning parameter, and the minimum value is considered over all the possible paths. For $\alpha <1$, a shorter path composed of weak connections is favored over a longer path with strong ties. On the contrary, when $\alpha >1$, the impact of additional intermediary nodes is relatively unimportant compared to the strength of the connection. So, paths with more intermediaries are favored. When $\alpha = 0$, the outcome is the shortest path distance of an undirected and unweighted graph, whereas when $\alpha = 1$, the outcome is the one obtained with Dijkstra's algorithm \cite{dijkstra59}. Note that with the previous method an infinite large distance would be assigned to absent ties (weight of 0).

Finally, it is worth mentioning that, according to Eq.~\ref{adj_transf_1}, all the elements of the $d-$path transformed adjacency matrix are less than or equal to one (i.e., $\hat{\mathbf{A}}^\tau(i,j) \leq 1$). Consequently, in case of $\alpha \geq 1$, the weighted shortest path distance between nodes $i$ and $j$, $d_{i,j}^\alpha$, is equal to the shortest path distance between those nodes when there are no LRIs (i.e., $d_{i,j}^\alpha=d_{i,j}$). 

%%%%%%%%%%%%%%%%%%%%%%%%%%%%%%%%%%%%%%%%%%%%
%%%%%%%%%%%%%%%%%%%%%%%%%%%%%%%%%%%%%%%%%%%%

\subsection{Estimation of MSD on weighted networks}

Given an initial condition $\vec{p}_{0,i}$, we find the MSD of the random walker to the origin (i.e., the node $i$), at each time step, $r^2(t,i)$. Let $ d_{i,j}^\alpha$ be the minimal distance from node $j$ to the origin, $i$. Then, $r^2(t,i)$ can be expressed as follows:

\begin{equation}
r^2(t,i)=\sum_{j=1}^{N} \left ( d_{i,j}^\alpha \right )^2 \left ( \vec{p}_{t,i} \right )_j.
\label{MSD_una_condicion}
\end{equation}

To estimate the MSD, we average over all the different initial positions of the walker:

\begin{equation}
\mathrm{MSD}\equiv \left \langle r^2(t) \right \rangle = \frac{1}{N}\sum_{i=1}^N r^2(t,i)= \frac{1}{N}\sum_{i=1}^N \sum_{j=1}^{N}\left ( d_{i,j}^\alpha \right )^2 \left ( \vec{p}_{t,i} \right )_j.
\label{eq_MSD_multpl}
\end{equation}

\noindent As can be observed, the value of MSD given by \ref{eq_MSD_multpl} is equal to Eq.~\ref{eq_MSD_multpl_new}, when $\alpha \geq 1$ (i.e., $d_{i,j}^\alpha=d_{i,j}$).

%%%%%%%%%%%%%%%%%%%%%%%%%%%%%%%%%%%%%%%%%%%%%%%%%%%%%%%%%%%

%%%%%%%%%%%%%%%%%%%%%%%%%%%%%%%%%%%%%%%%%%%%%%%%%%
\section*{APPENDIX C: Analytical expressions for MSD on cycle graphs with time-dependent LRIs.}
\label{Analytic_cicles}

Let $C=(V,E)$ be a simple, undirected cycle graph without self-loops and let the parameters of the Mellin and Laplace transformations be constant. Then, the transformed $d-$path adjacency matrices of $C$, $\hat{\mathbf{A}}^\tau$, are circulant matrices for every $t$ (see Appendix A).

Here are two immediate consequence of the above fact:

\begin{enumerate}[label=(\roman*)]
\item The strength of a given node of a transformed $d-$path graph $\hat{s}^\tau \left ( i \right )$ does not depend on the node $i$ (i.e., $\hat{s}^\tau \left ( i \right )=\hat{s}^\tau$).
\item The transition matrix for the random walk is symmetric:
\begin{equation}
{ \mathbf{\mathcal{P}}}={ \mathbf{\mathcal{P}}} ^T=\frac{1}{\hat{s}^\tau}\hat{\mathbf{A}}^\tau.
\end{equation}
\end{enumerate}

All circulant matrices are diagonalized in the Fourier basis:

\begin{equation}
\hat{\mathbf{A}}^\tau = \mathbf{U} \mathbf{D} \mathbf{U}^{\ast},
\end{equation}

\noindent where $\mathbf{U}$ is the unitary discrete Fourier transform matrix, $ \mathbf{U}^{\ast}$ is its conjugate transpose and $D$ is the diagonal matrix of eigenvalues of $\hat{\mathbf{A}}^\tau$. Consequently, Eq.~\ref{time_evolution_prob} can be expressed as follows:

\begin{equation}
\vec{p}_{t,i}= \left ( \prod_{\delta=0}^{t-1}{ \mathbf{\mathcal{P}}}_{\delta} ^T \right ) \vec{p}_{0,i}= \mathbf{U} \left ( \prod_{\delta=0}^{t-1}\frac{1}{\hat{s}^\tau}{ \mathbf{D}} \right )  \mathbf{U}^{\ast} \vec{p}_{0,i}=\mathbf{U} \mathbf{\mathcal{D}}_{t-1}  \mathbf{U}^{\ast} \vec{p}_{0,i},
\label{time_evolution_prob_cycle}
\end{equation}

\noindent where

\begin{equation}
\mathbf{\mathcal{D}}_{t-1} (i,i) = \prod_{\delta=0}^{t-1} \frac{\sigma_{i-1}^\tau}{\hat{s}^\tau}=\left ( \frac{\sigma_{i-1}^\tau}{\hat{s}^\tau} \right )^t,
\label{diagonal_prob_cycle}
\end{equation}

\noindent and $\sigma_{i-1}^\tau$ is the $i$th-eigenvalue of $\hat{\mathbf{A}}^\tau$.

If $N$ is an odd number and $\tau=\mathrm{Mellin}$, it is possible to write:

\begin{equation}
\sigma_{k}^\mathrm{Mellin}=\sum_{d=1}^{(N-1)/2} \frac{1}{d^{s}} \left ( \omega _k^{N-d} +\omega _k^{d} \right )=\sum_{d=1}^{(N-1)/2} \frac{2}{d^{s}}\cos\left ( \frac{2\pi kd}{N} \right ),
\label{autovalor_cyclo_finito1}
\end{equation}

and

\begin{equation}
\hat{s}^\mathrm{Mellin}=\sum_{d=1}^{(N-1)/2} \frac{2}{d^{s}}=2 H_{\frac{N-1}{2}} ^{(s)},
\label{strenght_cyclo_finito1}
\end{equation}

\noindent for $k=0,1,\cdots,N-1$, where $H_n^{(m)}$ is the generalized harmonic number of order $n$ of $m$,

\begin{equation}
\omega_k=\exp\left (\mathfrak{i} \frac{2\pi k}{N} \right ),
\end{equation}

\noindent are the $N-$th roots of unity, and $\mathfrak{i}=\sqrt{-1}$ (see Appendix A). Note that $\omega _k^{N}=1$ and, consequently, $\omega _k^{N-d}=\omega _k^{-d}$. As can be observed, it is straightforward to obtain a similar equation for $\tau=\mathrm{Laplace}$.

The unitary discrete Fourier transform matrix $\mathbf{U}$ can be defined as:

\begin{equation}
\mathbf{U}(i,j) =\left ( \frac{\omega_{-1}^{(i-1)(j-1)}}{\sqrt{N}} \right ).
\end{equation}

Consequently, according to Eq.~\ref{time_evolution_prob_cycle}, given an initial condition $\vec{p}_{0,i}$, the probability of finding a random walker at node $q$, at time $t$, is given by:

\begin{equation}
\left ( \vec{p}_{t,i} \right )_q = \sum_{j=1}^N \sum_{k=1}^N \mathbf{U}(q,k) \mathbf{\mathcal{D}}_{t-1} (k,k) \mathbf{U}^{\ast}(k,j) \left ( \vec{p}_{0,i} \right )_j.
\label{prob_nodo_ciclo}
\end{equation}

According to the topology of the cycle graphs, $r^2(t,i)=r^2(t,j)$ for $i \neq j$ (see Eq.~\ref{MSD_una_condicion}). Therefore, using Eq.~\ref{prob_nodo_ciclo}, if $N$ is odd, it is possible to write Eq.~\ref{eq_MSD_multpl} (see Appendix B) as follows:

\begin{eqnarray}
\left \langle r^2(t) \right \rangle = \frac{1}{N}\sum_{q=1}^N r^2(t,q)=  r^2(t,i)=\sum_{q=1}^{N}\left ( d_{q,i}^\alpha \right )^2 \left ( \vec{p}_{t,i} \right )_q=\nonumber \\
=\sum_{q=1}^N \sum_{j=1}^N \sum_{k=1}^N \left ( i-q \right )^2 \mathbf{U}(q,k) \mathbf{\mathcal{D}}_{t-1} (k,k) \mathbf{U}^{\ast}(k,j) \left ( \vec{p}_{0,i} \right )_j,
\label{eq_MSD_cycle}
\end{eqnarray}

\noindent for cycle graphs, when $i=\frac{N+1}{2}$ and $\alpha \geq 1$ (i.e., $\left ( d_{q,i}^\alpha \right )^2=\left ( d_{q,i} \right )^2=\left ( i-q \right )^2$). According to the definition of the vector $\vec{p}_{0,i}$, Eq.~\ref{eq_MSD_cycle} results in:

\begin{eqnarray}
\left \langle r^2(t) \right \rangle
=\sum_{q=1}^N \sum_{k=1}^N \left ( i-q \right )^2  \frac{1}{N} \exp\left ( \mathfrak{i}\theta_k \left ( i-q \right ) \right ) \mathbf{\mathcal{D}}_{t-1} (k,k) =\nonumber \\
= \sum_{q=1}^N \left ( i-q \right )^2  \frac{1}{N} +\sum_{q=1}^N \sum_{k=2}^N \left ( i-q \right )^2  \frac{1}{N} \exp\left ( \mathfrak{i} \theta_k \left ( i-q \right ) \right ) \mathbf{\mathcal{D}}_{t-1} (k,k) =\nonumber \\
=\frac{N^2-1}{12} + \sum_{k=2}^N \frac{1}{N} \mathbf{\mathcal{D}}_{t-1} (k,k)  \sum_{q=1}^N \left ( i-q \right )^2 \exp\left ( \mathfrak{i} \theta_k \left ( i-q \right ) \right )=\nonumber \\
=\frac{N^2-1}{12} +\sum_{k=2}^N  \mathbf{\mathcal{D}}_{t-1} (k,k) \frac{\exp\left ( \mathfrak{i} \theta_k i\right )}{\exp\left ( \mathfrak{i} \theta_k \right )-1}\left ( 2(i-1)+N-\frac{2}{\exp\left ( \mathfrak{i} \theta_k\right )-1} \right )=\nonumber \\
=\frac{N^2-1}{12} +\sum_{k=2}^N  \mathbf{\mathcal{D}}_{t-1} (k,k)\Phi (k)
\label{eq_MSD_cycle_final_2},
\end{eqnarray}

\noindent where we introduce the short-hand notation

\begin{equation}
\Phi (k) \equiv \frac{\exp\left ( \mathfrak{i} \theta_k i\right )}{\exp\left ( \mathfrak{i} \theta_k \right )-1}\left ( 2(i-1)+N-\frac{2}{\exp\left ( \mathfrak{i} \theta_k\right )-1} \right )=\frac{(-1)^k}{2\sin \left ( \frac{\theta _k}{2} \right )}
\left ( 2N\mathfrak{i} - \cot\left ( \frac{\theta _k}{2} \right )\right ),
\label{phi_def}
\end{equation}

\noindent and $\theta_k \equiv \frac{2\pi}{N}\left ( k-1 \right )$, for $i=\frac{N+1}{2}$ (see Secs. D and F of Appendix B). As can be observed, Eq.~\ref{eq_MSD_cycle_final_2} includes a constant term. It can be proved that it is the saturation value of MSD on cycle graphs (see Sec. E of Appendix B).

Given that $\theta_k \equiv -\theta_{N+2-k}$ (mod $2\pi$), $\mathbf{\mathcal{D}}_{t-1} (k,k)=\mathbf{\mathcal{D}}_{t-1} (N+2-k,N+2-k)$ and $\Phi (k) = \Phi ^* (N+2-k)$. Therefore, MSD can be expressed as:

\begin{eqnarray}
\left \langle r^2(t) \right \rangle= \frac{N^2-1}{12} +\sum_{k=2}^{(N+1)/2}  \mathbf{\mathcal{D}}_{t-1} (k,k)\frac{(-1)^{k+1}}{\sin \left ( \frac{\theta _k}{2} \right )} \cot\left ( \frac{\theta _k}{2} \right ),
\label{eq_MSD_cycle_final_3}
\end{eqnarray}

\noindent on cycle graphs with an odd number of nodes.

%%%%%%%%%%%%%%%%%%%%%%%%%%%%%%%%%%%%%%%%%%%%%%%%%%%%%%

\subsection{Time evolution of the MSD on cycle graphs without LRIs}

In case of $s=\infty$ for every $t$, $\mathbf{\mathcal{D}}_{t-1} (k,k)= \cos^t\left ( \theta_k \right )$. Thus, according to Eq.~\ref{eq_MSD_cycle_final_3}, MSD is given by:

\begin{eqnarray}
\left \langle r^2(t) \right \rangle= \frac{N^2-1}{12} +\sum_{k=2}^{(N+1)/2}  \cos^t\left ( \theta_k \right )\frac{(-1)^{k+1}}{\sin \left ( \frac{\theta _k}{2} \right )} \cot\left ( \frac{\theta _k}{2} \right ).
\label{eq_MSD_s_inf}
\end{eqnarray}

%%%%%%%%%%%%%%%%%%%%%%%%%%%%%%%%%%%%%%%%%%%%%%%%%%%%
\subsection{Time evolution of the MSD on cycle graphs with LRIs}

In case of $s=0$ for every $t$, we obtain the following result:

\begin{equation}
\mathbf{\mathcal{D}}_{t-1} (k,k) =  \left ( \frac{2}{N-1}\csc\left ( \frac{\theta _t}{2}\right )\sin\left ( \frac{\theta _t\left (N-1\right )}{4} \right ) \cos\left ( \frac{\theta _t\left (N+1\right )}{4}   \right ) \right ) ^t.
\end{equation}

%Generally, in case of constant $s$, taking into account that $d/N < 0.5$, Eq.~\ref{autovalor_cyclo_finito1} can be written as follows:

%\begin{eqnarray}
%\sigma_{k,\delta}^\mathrm{Mellin}=\sum_{d=1}^{(N-1)/2} \frac{2}{d^{s}}\cos\left ( \frac{2\pi kd}{N} \right )\approx
%\sum_{d=1}^{(N-1)/2} \frac{2}{d^{s}} \left ( 1+\frac{1}{2}\left ( \frac{2\pi kd}{N} \right )^2 \right )=\nonumber\\=
%2 H_{\frac{N-1}{2}} ^{(s)}+\left ( \frac{2\pi k}{N} \right )^2 H_{\frac{N-1}{2}} ^{(s-2)}.
%\end{eqnarray}

%Thus, Eqs. \ref{diagonal_prob_cycle} and \ref{eq_MSD_cycle_final_3} can be expressed as

According to Eqs. \ref{autovalor_cyclo_finito1} and \ref{strenght_cyclo_finito1}, in case of constant $s$, Eq.~\ref{diagonal_prob_cycle} leads to:

\begin{equation}
\mathbf{\mathcal{D}}_{t-1} (k,k) =\prod_{\delta=0}^{t-1}  \frac{\sigma_{k-1}^\tau}{\hat{s}^\tau} = \left ( \frac{1}{H_{0,\frac{N-1}{2}} ^{s}} \sum_{d=1}^{(N-1)/2}\frac{\cos\left ( \theta_k d\right )}{d^{s}} \right )^t.
\label{term_temp_s_cons}
\end{equation}

\noindent Therefore, replacing Eq.~\ref{term_temp_s_cons} into Eq.~\ref{eq_MSD_cycle_final_3}, we obtain Eq.~\ref{eq_MSD_s_cons}.

%%%%%%%%%%%%%%%%%%%%%%%%%%%%5
%%%%%%%%%%%%%%%%%%%%%%%%%%%%%%%%%%%%%%%%%%%%%%%%%%%%%%

%%%%%%%%%%%%%%%%%%%%%%%%%%%%%%%%%%%%%%%%%%%%%%%%
\subsection{Operations on Eq.~\ref{eq_MSD_cycle_final_2}.}

To simplify the right hand term of Eq.~\ref{eq_MSD_cycle_final_2}, we proceed as follows:

\begin{eqnarray}
 \sum_{k=2}^N \frac{1}{N} \mathbf{\mathcal{D}}_{t-1} (k,k)  \sum_{q=1}^N \left ( i-q \right )^2 \exp\left ( \mathfrak{i} \theta_k \left ( i-q \right ) \right )=\nonumber \\
\sum_{k=2}^N \frac{1}{N} \mathbf{\mathcal{D}}_{t-1} (k,k) \frac{1}{\left ( \mathfrak{i} \theta_k \right )^2} \sum_{q=1}^N \left ( \mathfrak{i} \theta_k \left ( i-q \right ) \right )^2 \exp\left ( \mathfrak{i} \theta_k \left ( i-q \right ) \right ).
\label{semi_step}
\end{eqnarray}

The summation on $q$ of Eq.~\ref{semi_step} results in:

\begin{eqnarray}
 \sum_{q=1}^N \left ( \mathfrak{i} \theta_k \left ( i-q \right ) \right )^2 \exp\left ( \mathfrak{i} \theta_k \left ( i-q \right ) \right )
=\sum_{q=1} ^{N} \frac{\partial ^2}{\partial a^2} \left [  \exp\left ( \mathfrak{i} \theta_k \left ( i-q \right )a \right ) \right ]_{a=1}= \nonumber\\
=\frac{\partial ^2}{\partial a^2}\left [ \exp\left (\mathfrak{i} \theta_k i a \right ) \sum _{q=1}^{N}\left ( \exp\left ( -\mathfrak{i} \theta_k  a \right ) \right )^q\right ]_{a=1}=\nonumber\\
=\frac{\partial ^2}{\partial a^2}\left [ \exp\left (\mathfrak{i} \theta_k (i-1) a \right ) \frac{1-\exp\left ( -\mathfrak{i} \theta_k  a N \right )}{1-\exp\left ( -\mathfrak{i} \theta_k  a \right )} \right ]_{a=1}=\nonumber\\
=N\left ( \mathfrak{i} \theta_k  \right )^2\exp\left ( \mathfrak{i} \theta_k  i\right )
\left ( 
\frac{\left ( i-1 \right )^2 \left ( 1-\exp\left ( -\mathfrak{i} \theta_k N\right ) \right ) }{N\left ( \exp\left ( \mathfrak{i} \theta_k \right ) -1 \right )}-
\frac{2\left ( i-1 \right ) \left ( 1-\exp\left ( -\mathfrak{i} \theta_k N\right ) \right ) }{N\left ( \exp\left ( \mathfrak{i} \theta_k \right ) -1 \right )^2}+\right . \nonumber\\
+\frac{2\left ( i-1 \right ) \exp\left ( -\mathfrak{i} \theta_k N\right ) }{\exp\left ( \mathfrak{i} \theta_k \right ) -1}+
\frac{N \exp\left ( -\mathfrak{i} \theta_k N\right ) }{\exp\left ( \mathfrak{i} \theta_k \right ) -1}+\nonumber\\
\left .
+\frac{\left (\exp\left ( \mathfrak{i} \theta_k \right )+1 \right ) \left ( 1-\exp\left ( -\mathfrak{i} \theta_k N\right ) \right ) }{N\left ( \exp\left ( \mathfrak{i} \theta_k \right ) -1 \right )^3}-
\frac{2 \exp\left ( -\mathfrak{i} \theta_k N\right ) }{\left ( \exp\left ( \mathfrak{i} \theta_k \right ) -1 \right )^2}
 \right ).
\nonumber\\
\end{eqnarray}

Finally, taking into account that $\exp\left ( \mathfrak{i} \theta_k N\right )=\exp\left ( -\mathfrak{i} \theta_k N\right )= 1$, we obtain the result presented in Eq.~\ref{eq_MSD_cycle_final_2}.

\subsection{Saturation value of MSD on cycle graphs}

If $k\geq2$ and $s(t)>0$, according to Eqs. \ref{autovalor_cyclo_finito1} and \ref{strenght_cyclo_finito1}, $\sigma_{k-1}^\mathrm{Mellin}<\hat{s}^\mathrm{Mellin}$. Let $r$ be equal to $\sigma_{k-1}^\mathrm{Mellin} / \hat{s}^\mathrm{Mellin}$. Then, it is possible to proof that the following limit tend to zero:

\begin{eqnarray}
\lim_{t\rightarrow \infty}
\sum_{k=2}^N  \mathbf{\mathcal{D}}_{t-1} (k,k) \frac{\exp\left ( \mathfrak{i} \theta_k i\right )}{\exp\left ( \mathfrak{i} \theta_k \right )-1}\left ( 2(i-1)+N-\frac{2}{\exp\left ( \mathfrak{i} \theta_k\right )-1} \right )=\nonumber \\
\sum_{k=2}^N
\left (\lim_{t\rightarrow \infty} \prod_{\delta=0}^{t-1} \frac{\sigma_{k-1}^\mathrm{Mellin}}{\hat{s}^\mathrm{Mellin}}\right )\frac{\exp\left ( \mathfrak{i} \theta_k i\right )}{\exp\left ( \mathfrak{i} \theta_k \right )-1}\left ( 2(i-1)+N-\frac{2}{\exp\left ( \mathfrak{i} \theta_k\right )-1} \right ) = \nonumber \\
=\sum_{k=2}^N
\left (\lim_{t\rightarrow \infty} r^t \right )\frac{\exp\left ( \mathfrak{i} \theta_k i\right )}{\exp\left ( \mathfrak{i} \theta_k \right )-1}\left ( 2(i-1)+N-\frac{2}{\exp\left ( \mathfrak{i} \theta_k\right )-1} \right )=0. 
\label{saturation_limit}
\end{eqnarray} 

Consequently,  according to Eq.~\ref{eq_MSD_cycle_final_2}, we obtain:

\begin{eqnarray}
\lim_{t\rightarrow \infty}\left \langle r^2(t) \right \rangle = \frac{N^2-1}{12}.
\end{eqnarray} 

%%%%%%%%%%%%%%%%%%%%%%%%%%%%%%%%%%%%%%%%%%%%%%%%%%%%%

\subsection{On $\Phi (k)$}

According to the definition of $\Phi (k)$ (Eq.~\ref{phi_def}), for $i=\frac{N+1}{2}$, it is possible to write:

\begin{eqnarray} 
\Phi (k) \equiv \frac{\exp\left ( \mathfrak{i} \theta_k i\right )}{\exp\left ( \mathfrak{i} \theta_k \right )-1}\left ( 2(i-1)+N-\frac{2}{\exp\left ( \mathfrak{i} \theta_k\right )-1} \right )=\nonumber \\
=\exp\left ( \mathfrak{i} \theta_k i\right )\frac{\exp\left ( -\mathfrak{i} \theta_k \right )-1}{2\left ( 1-\cos\left ( \theta _k \right ) \right )}\left ( 2N-1-2 \frac{\exp\left ( -\mathfrak{i} \theta_k \right )-1}{2\left ( 1-\cos\left ( \theta _k \right ) \right )}\right )=\nonumber \\
=\frac{\exp\left ( \mathfrak{i} \theta_k \frac{N-1}{2}\right )-\exp\left ( \mathfrak{i} \theta_k \frac{N+1}{2}\right )}{2\left ( 1-\cos\left ( \theta _k \right ) \right )}\left ( 2N-1-2 \frac{\cos\left ( \theta _k \right ) -1-\mathfrak{i}\sin\left ( \theta _k \right )}{2\left ( 1-\cos\left ( \theta _k \right ) \right )}\right )=\nonumber \\
=\frac{\exp\left ( \mathfrak{i} \theta_k \frac{N}{2}\right )(-\mathfrak{i})\sin\left ( \frac{\theta _k}{2} \right )}{2\sin^2\left ( \frac{\theta _k}{2} \right )}\left ( 2N+ \frac{\mathfrak{i}2\sin\left ( \frac{\theta _k}{2} \right )\cos\left ( \frac{\theta _k}{2} \right )}{2\sin^2\left ( \frac{\theta _k}{2} \right )}\right )=\nonumber \\
=\frac{\exp\left ( \mathfrak{i} \pi k\right )\exp\left ( -\mathfrak{i} \pi\right )(-\mathfrak{i})}{2\sin \left ( \frac{\theta _k}{2} \right )}
\left ( 2N+\mathfrak{i} \cot\left ( \frac{\theta _k}{2} \right )\right )=\nonumber \\
=\frac{(-1)^k}{2\sin \left ( \frac{\theta _k}{2} \right )}
\left ( 2N\mathfrak{i} - \cot\left ( \frac{\theta _k}{2} \right )\right ).
\end{eqnarray}

\end{document}